\documentclass[fleqn,usenatbib]{mnras}
\usepackage{newtxtext,newtxmath}
\usepackage[T1]{fontenc}
\DeclareRobustCommand{\VAN}[3]{#2}
\let\VANthebibliography\thebibliography
\def\thebibliography{\DeclareRobustCommand{\VAN}[3]{##3}\VANthebibliography}

\usepackage{graphicx}
\usepackage{amsmath}
\usepackage{natbib}
\usepackage{xspace}
\usepackage{bm}
\usepackage{booktabs}
\usepackage{tabularx}
\usepackage{xcolor}
\usepackage{listings}
\usepackage{multirow}
\usepackage{hyperref}

\newcommand{\package}{\texttt{spright}\xspace}
\newcommand{\rearth}{\ensuremath{R_\oplus}\xspace}
\newcommand{\gcm}{\ensuremath{\mathrm{g\,cm^{-3}}}\xspace}
\newcommand{\emcee}{\texttt{emcee}\xspace}
\renewcommand{\vec}[1]{\pmb{#1}}

\newcommand{\tess}{\textit{TESS}\xspace}

\newcommand{\pww}{\ensuremath{\omega}\xspace}
\newcommand{\pwc}{\ensuremath{\psi}\xspace}
\newcommand{\catsize}{\ensuremath{N_\mathrm{p}}\xspace}
\newcommand{\bfactor}{\ensuremath{2 \ln B}\xspace}

\title[\texttt{spright}]{\texttt{Spright}: a probabilistic mass-density-radius relation for small planets}

\author[H. Parviainen et al.]{
Hannu Parviainen,$^{1,2}$\thanks{E-mail: hannu@iac.es (HP)}
Rafael Luque,$^{3}$
Enric Palle$^{2,1}$
\\
$^{1}$Dept. Astrof\'isica, Universidad de La Laguna (ULL), E-38206 La Laguna, Tenerife, Spain \\
$^{2}$Instituto de Astrof\'isica de Canarias (IAC), E-38200 La Laguna, Tenerife, Spain\\
$^{3}$Department of Astronomy \& Astrophysics, University of Chicago, Chicago, IL 60637, USA
}

\date{Accepted XXX. Received YYY; in original form ZZZ}
\pubyear{2023}

\begin{document}
\label{firstpage}
\pagerange{\pageref{firstpage}--\pageref{lastpage}}
\maketitle

\begin{abstract}
We present \package, a \texttt{Python} package that implements a fast and lightweight mass-density-radius relation for small planets. The relation represents the joint planetary radius and bulk density probability distribution as a mean posterior predictive distribution of an analytical three-component mixture model. The analytical model, in turn, represents the probability for the planetary bulk density as three generalised Student’s t-distributions with radius-dependent weights and means based on theoretical composition models. The approach is based on Bayesian inference and aims to overcome the rigidity of simple parametric mass-radius relations and the danger of overfitting of non-parametric mass-radius relations.

The package includes a set of pre-trained and ready-to-use relations based on two M dwarf catalogues, one FGK star catalogue, and two theoretical composition models for water-rich planets. The inference of new models is easy and fast, and the package includes a command line tool that allows for coding-free use of the relation, including the creation of publication-quality plots.

Additionally, we study whether the current mass and radius observations of small exoplanets support the presence of a population of water-rich planets positioned between rocky planets and sub-Neptunes. The study is based on Bayesian model comparison and shows somewhat strong support against the existence of a water-world population around M dwarfs. However, the results of the study depend on the chosen theoretical water-world density model. A more conclusive result requires a larger sample of precisely characterised planets and community consensus on a realistic water world interior structure and atmospheric composition model.
\end{abstract}


\section{Introduction}
\label{sec:introduction}

The launch of the Transiting Exoplanet Survey Satellite \citep[\tess;][]{TESS} has enabled the bulk density measurement of hundreds of exoplanets thanks to its all-sky observing strategy of nearby, bright stars. Its contribution is already comparable to that of the \textit{Kepler/K2} mission for small planets ($R < 4\,\rearth$) with mass determinations. Out of the thousands of small exoplanets discovered by \textit{Kepler/K2}, only a few hundred have bulk density measurements ($\sim 330$, based on the NASA Exoplanet Archive\footnote{\url{https://exoplanetarchive.ipac.caltech.edu/index.html}} as of May 2023). On the other hand, approximately 130 small planets discovered by \tess\ have precisely determined masses and radii to date, with hundreds of candidates awaiting to be confirmed and characterised with ground-based facilities. For M dwarfs, the contribution is even larger --- approximately 7\% of all the \textit{Kepler/K2} small planets with measured bulk densities orbit M-dwarf hosts, while for \tess, the ratio is 40\%. However, such in-depth characterisation is observationally expensive and it becomes harder as the planet-host mass and size ratios decrease.

Probabilistic mass-radius (M-R) relationships are useful not only for the purpose of predicting one quantity from the other but also as a means of understanding exoplanet compositions. On the one hand, they allow feasibility studies and efficient planning of radial velocity (RV) and transmission spectroscopy observations of transiting planets, which require an estimate of the mass given a radius measurement to predict the expected RV semi-amplitude and the planet's atmosphere scale height (which is inversely proportional to the planet's gravity and relates to its detectability), respectively. Reversely, upcoming microlensing discoveries with, e.g., the \textit{Roman Space Telescope} \citep{Spergel2015} will have mass estimates for which a direct radius measurement is impossible. 

On the other hand, M-R relationships are a robust tool to identify demographic features of the exoplanet population, such as the transition from brown dwarfs to hydrogen-burning stars \citep[e.g.,][]{HatzesRauer15}, Neptunian to Jovian planets \citep[e.g.,][]{Wolfgang2016, ChenKipping17, Bashi17}, and rocky to volatile-rich planets \citep[e.g.,][]{Weiss2014, Zeng2019, Otegi20, Luque2022}. Linking these trends with the physical and chemical processes at play during planet formation and evolution offers an avenue to constrain observationally such theories. 

This is especially relevant for sub-Neptune-sized exoplanets --- with radii between 1.5 and 4\,$R_\oplus$ --- whose nature and origin are actively debated. Unlikely to be rocky in nature \citep{Rogers15, Fulton17}, these planets reside in a degenerate part of the M-R parameter space where their bulk densities are equally well explained by solid rocky/iron cores with primordial gaseous hydrogen-rich atmospheres (sometimes referred to as "gas dwarfs", \citealt{Lopez2014, Rogers2023}) or a water-rich interior and atmosphere akin to the icy moons of the solar system (sometimes referred to as "water worlds", \citealt{Leger2004, DornLichtenberg2021, Aguichine21}). Both gas dwarfs and water worlds are naturally explained by current planet formation and evolution models \citep[see e.g.,][]{LeeChiang2016, OwenWu2017, Ginzburg2018, Bitsch2019, Venturini2020, Burn21}, but with remarkably different implications about their location at birth. While the prevailing view is that sub-Neptunes are primarily gas dwarfs \citep[see the review by][]{Bean2021}, the existence of water worlds appears strongly supported by recent individual planet discoveries \citep{Bluhm2021, DiamondLowe2022, Piaulet2023, Acuña2022}, observational demographic studies \citep{Zeng2019, Neil2022, Luque2022}, and advances in interior structure and global formation modeling \citep{Venturini2020, Burn21, DornLichtenberg2021, Aguichine21}.

Most of the previous studies on M-R relations have assumed that exoplanet masses and radii follow one or multiple power-law segments of the form $M \propto R^\gamma$ \citep{Lissauer11, Weiss2014, Wolfgang2016, Mills2017, ChenKipping17, Bashi17, Otegi20}. Others have proposed a non-parametric approach instead \citep{Ning2018, Kanodia2019}. The power-law models are simple to fit, and their parameters are easy to interpret, but their rigidity also means that they may give an overly simplistic representation of the actual M-R relation. On the contrary, non-parametric models can take on a variety of shapes to fit the data and do not assume the distribution of masses at a given radius to be Gaussian or even symmetric, but their flexibility plays against their precision for small sample sizes. 

In this paper, we present \package,\!\footnote{\url{https://github.com/hpparvi/spright}, \href{https://doi.org/10.5281/zenodo.10082653}{DOI:10.5281/zenodo.10082653 }} a \texttt{Python} package that provides a lightweight probabilistic M-R relation for small planets.
The relation models the joint planetary radius and bulk density distribution as a mean of the posterior predictive distribution of a simple analytic three-component mixture model. The approach is based on basic Bayesian inference and 
aims to overcome the shortcomings of the existing methods by 1) delivering robustness and flexibility not offered by parametric models while 2) avoiding the dangers of overfitting and the need for large sample sizes associated with non-parametric approaches. We detail how the model is constructed, compare our results with previous M-R relations, and study how the posterior radius-density model compares going from M-dwarf to solar-type hosts. Finally, we also explore the use of \package to study whether the current observations support the existence of water worlds as a separate population between rocky planets and sub-Neptunes, as suggested by \citet{Luque2022}. 

\section{Methods}
\label{sec:methods}

\subsection{Overview}
\label{sec:methods.overview}

The \package package provides a numerical radius-density-mass relation (referred to as "numerical model" from now on) for small planets. The numerical model is constructed by averaging an analytical three-component mixture probability model ("analytical model", from now on) over its posterior parameter space given a catalogue of empirical planet mass and radius measurements. The analytical model is composed of three generalised Student's t-distributions with the distribution means and weights varying as functions of the planetary radius. The three components correspond to rocky planets, water-rich planets (water worlds), and hydrogen-rich sub-Neptunes, and the model parameterisation is designed so that the water world component is optional. That is, the analytical model can represent the observed mass-radius distribution either as a mixture of rocky planets and sub-Neptunes, or as a mixture of rocky planets, water-rich planets, and sub-Neptunes. Thus, the final numerical average model is agnostic to whether water worlds exist as a distinct population.

\subsection{Analytical radius-density probability model}
\label{sec:methods.analytical_model}
\subsubsection{Probability distribution}
\label{sec:methods.analytical_model.distribution}

The analytical radius-density relation implemented in \package models the probability distribution for a planet's bulk density, $\rho$, given the planet's radius, $r$, and model parameter vector, $\vec{\theta}$, as a mixture of three generalised (scaled and transformed) Student's t-distributions with five degrees of freedom,\!\footnote{We chose the Student's t-distribution with five degrees of freedom ($\lambda=5$) instead of a normal distribution because the t-distribution's heavier tails make the inference less sensitive to outliers. We first tested treating $\lambda$ as a free parameter, but this complicates fitting to observations since $\lambda$ is degenerate with the distribution's scale parameter. Ultimately, we chose $\lambda=5$ because it leads to a numerically cheap analytical probability distribution function (pdf) and provides some robustness over the normal distribution.} as 
\begin{equation}
P(\rho|r,\vec{\theta}) = w_\mathrm{r}(r) P_\mathrm{r}(r,\vec{\theta}_\mathrm{r}) + w_\mathrm{w}(r) P_\mathrm{w}(r,\vec{\theta}_\mathrm{w}) +  w_\mathrm{p}(r) P_\mathrm{p}(r,\vec{\theta}_\mathrm{p}).
\end{equation}
The distributions represent rocky planets ($P_\mathrm{r}$),  water-rich planets ($P_\mathrm{w}$), and hydrogen-rich sub-Neptunes ($P_\mathrm{p}$);
$w_\mathrm{r}$, $w_\mathrm{w}$, and $w_\mathrm{p}$ are the mixture weights with $w_\mathrm{r} + w_\mathrm{w} + w_\mathrm{p} = 1$ for all $r$; and $\vec{\theta}_\mathrm{r}$, $\vec{\theta}_\mathrm{w}$, and $\vec{\theta}_\mathrm{p}$ are the distribution-specific parameter vectors. 
More precisely, the distributions are
\begin{gather}
    P_\mathrm{r}(\rho|r, a, s_\mathrm{r}, \lambda_\mathrm{r}) = T(\mu = \rho_\mathrm{r}(a, r), s_\mathrm{r}, \lambda_\mathrm{r} = 5),\\
    P_\mathrm{w}(\rho|r, b, s_\mathrm{w}, \lambda_\mathrm{w}) = T(\mu = \rho_\mathrm{w}(b, r), s_\mathrm{w}, \lambda_\mathrm{w} = 5),\\
    P_\mathrm{p}(\rho|r, c, d, s_\mathrm{p}, \lambda_\mathrm{p}) = T(\mu = c r^d / 2^d, s_\mathrm{p}, \lambda_\mathrm{p} = 5),
\end{gather}
where $\mu$ are the distribution means, $s$ are the distribution scale parameters, $\lambda$ are the degrees of freedom, and $\rho_\mathrm{r}$ and $\rho_\mathrm{w}$ are the mean functions for rocky and water-rich planets, respectively. 

The package supports two theoretical radius-density models to represent the distribution means: the rocky-planet distribution mean follows the models by \citet[Z19, ][]{Zeng2019}\footnote{\url{https://lweb.cfa.harvard.edu/~lzeng/planetmodels.html}} and is parameterised by the iron-rock mixing ratio, $a$; the water-rich planet distribution mean follows the water-rich models either by \citet[A21, ][]{Aguichine21}\footnote{The models presented here have been calculated with the A21 models that assume $T_{\rm irr}=500$\,K and an Earth-like composition for the core. The full A21 grid will be made available in future.} or \citet{Zeng2019} and is parameterised by the H$_2$O-rock mixing ratio, $b$; and the sub-Neptune distribution mean is modelled as a power law with density at $2\rearth$ defined by $c$ and exponent by $d$. The choice to model the sub-Neptune population mean as a power law is motivated by the discussion in \citet{Lopez2014}.

\begin{figure}
    \centering
    \includegraphics[width=\columnwidth]{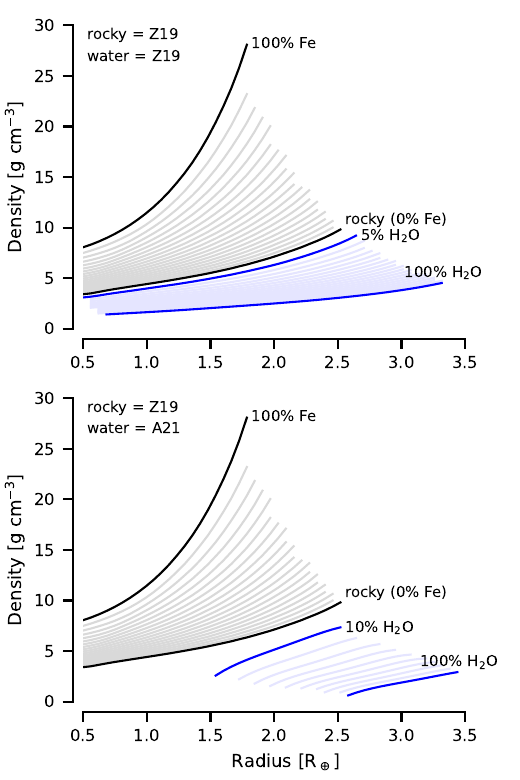}
    \caption{Theoretical models for the bulk planet density for rocky and water-rich planets by \citet[Z19, ][]{Zeng2019} and \citet[A21, ][]{Aguichine21}. The black lines show the models for rocky planets with iron-rock mixing ratio varying from 0\% to 100\%, and the blue lines show the water-rich planet models with H$_2$O-rock mixing ratio varying from 5\% to 100\%. The analytical radius-density probability model creates the rocky and water-rich planet distribution mean functions by interpolating inside the theoretical models.}
    \label{fig:density_mean_functions}
\end{figure}

\subsubsection{Mixture weights}
\label{sec:methods.analytical_model.weights}

\begin{figure*}
    \centering
    \includegraphics[width=\textwidth]{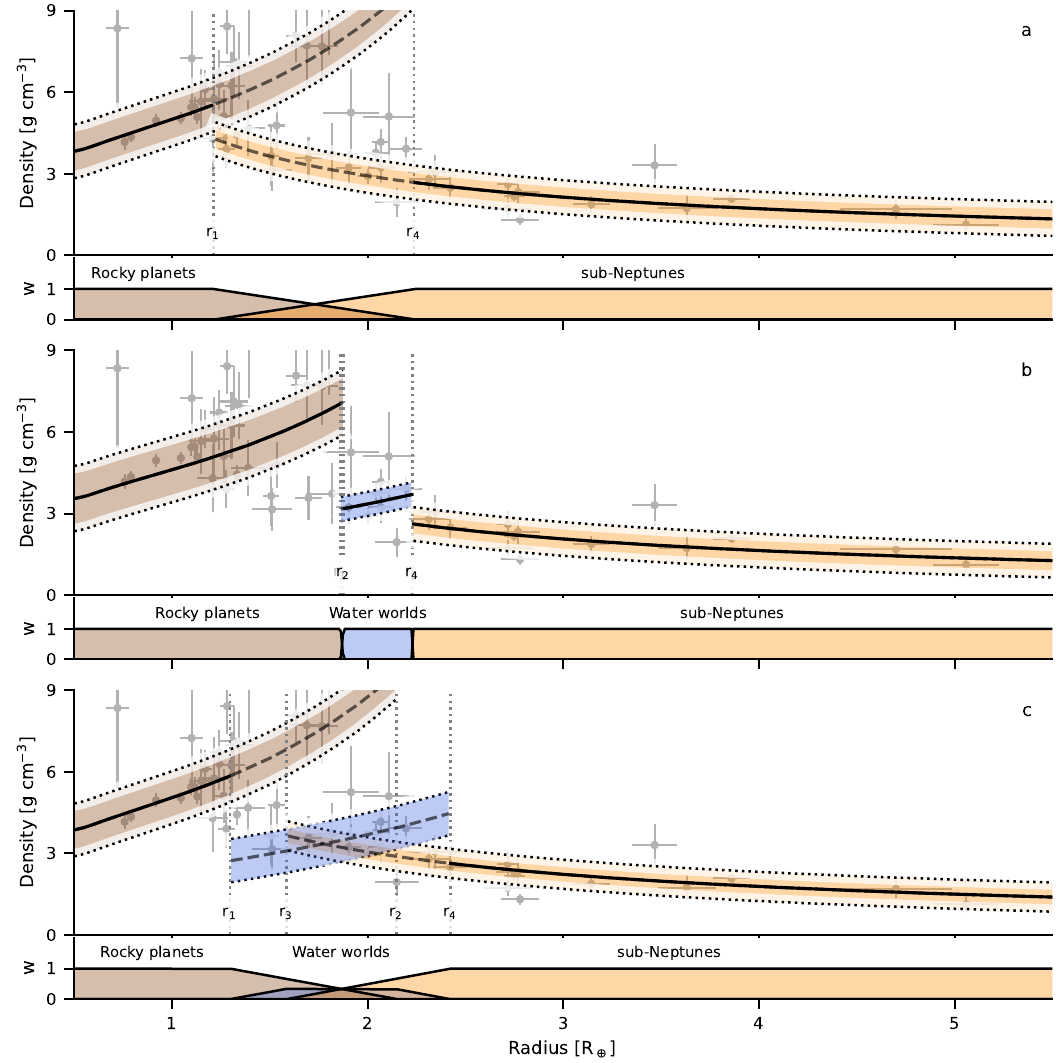}
    \caption{Three realisations of the analytical density mixture probability model used to calculate the final numerical radius-density probability model. The empirical dataset used to fit the model is the STPM catalogue used in Section~\ref{sec:discussion.catalogue_comparison}. The model allows for three distinct small-planet populations: rocky planets (brown), water worlds (blue), and hydrogen-rich sub-Neptunes (orange). The density probability is a mixture of three generalised Student's t-distributions with five degrees of freedom, mean following theoretical radius-density models by \citet{Aguichine21} or \citet{Zeng2019}, and width being a free parameter in the fit. The mixture weights ($w_\mathrm{r}$, $w_\mathrm{w}$, and $w_\mathrm{p}$) are calculated based on two transition regimes defined by ($r_1$ and $r_2$) and ($r_3$ and $r_4$). In each of the three panels, the upper sub-panel shows the mean and one-sigma limits for the density distributions for each mixture component: the solid line shows the radius regime where the component explains the density fully (without the need for other components), and the dashed line shows an overlapping region between components. The lower sub-panels show the weights for the mixture components. The model can explain the observed radius-density distribution with or without the water world population. Panel a) shows a model realisation with $\pww = 0$ (that is, $r_1 = r_3$ and $r_2 = r_4$ using Eqs.~\ref{eq:rtwo} and \ref{eq:rthree}), which leads to a direct transition from rocky planets to sub-Neptunes and excludes water worlds. Panel b) shows a realisation with $\pww = 1$ ($r_1 = r_2$ and $r_3 = r_4$), where water worlds are explained as a well-defined separate population between rocky planets and sub-Neptunes. Finally, panel c) shows a realisation with $\pww = 0.25$ corresponding to a weak water world component. Further details about the component weights can be found from a Jupyter notebook at \url{https://github.com/hpparvi/spright/blob/main/notebooks/A1_analytical_model_weights.ipynb}.}
    \label{fig:mixture_model}
\end{figure*}

The distribution weights vary as a function of the planetary radius, as shown in Fig.~\ref{fig:mixture_model}. The model divides the planet radius-space into three regimes parameterised by the rocky-water transition start and end radii, $r_1$ and $r_2$, and water-sub-Neptune transition start and end radii, $r_3$ and $r_4$. More precisely,
\begin{description}
    \item[$r_1$] is the radius limit below which all planets are rocky,
    \item[$r_2$] is the maximum radius for a rocky planet,
    \item[$r_3$] is the minimum radius for a puffy sub-Neptune, and
    \item[$r_4$] is the radius limit above which all planets are sub-Neptunes. 
\end{description}
The mixture weights are calculated by first mapping the planet's radius to a 2D triangle defined by the three composition classes, with rocky planets located at (0,0), water-rich planets at (1,0), and sub-Neptunes at (0,1).  The $(x,y)$ coordinates for any $r$ are
\begin{align}
x &= {\Bigl\lfloor} \frac{r-r_3}{r_4-r_3} {\Bigr\rceil},\\
y &= {\Bigl\lfloor} {\Bigl\lfloor} \frac{r-r_1}{r_2-r_1} {\Bigr\rceil} - x {\Bigr\rceil},
\end{align}
where ${\bigl\lfloor} v {\bigr\rceil}$ denotes clipping the the value $v$ between $0$ and $1$, that is, ${\bigl\lfloor} v {\bigr\rceil} = \max(0, \min(v, 1))$.
Next, the $(x,y)$ coordinates are mapped to the mixture weights through linear interpolation inside the composition triangle,\!\footnote{\url{https://en.wikipedia.org/wiki/Barycentric_coordinate_system}} as
\begin{gather}
    w_\mathrm{r} = 1 - x - y,\\
    w_\mathrm{w} = y,\\
    w_\mathrm{p} = 1 - w_\mathrm{r} - w_\mathrm{w},
\end{gather}
where the weight calculation is simplified from the general case due to the choice of the vertex locations.

\subsubsection{Parameterisation}
\label{sec:methods.analytical_model.parameterisation}

\begin{table}
\centering
\caption{Analytical mixture model parameters and priors. N$(\mu, \sigma)$ stands for a normal prior with a mean $\mu$ and standard deviation $\sigma$, and U$(a,b)$ stands for a uniform distribution from $a$ to $b$.}
\label{table:model_parameters}
\begin{tabular*}{\columnwidth}{@{\extracolsep{\fill}} llll}
\toprule\toprule
Description & Name & Units & Prior \\
\midrule     
Rocky planet transition start & $r_1$ & \rearth & U(0.5, 2.5) \\
Sub-Neptune transition end   & $r_4$ & \rearth & U(1.0, 4.0) \\
Water world population strength & \pww & - & U(0.0, 1.0) \\
Water world population shape  & \pwc & - & U(-1.0, 1.0) \\
Rocky planet iron ratio & $a$ & - & U(0.0, 1.0)\\
Water-rich planet water ratio & $b$ & - & N(0.5, 0.1)\\
Sub-Neptune density at $r=2$ & $c$ & \gcm & U(0.1, 7.0)\\
Sub-Neptune density exponent & $d$ & & N(-0.5, 1.5)\\
Log$_{10}$ rocky planet pdf scale & $s_r$ & $\log_{10}$ \gcm & N(0.0, 0.35)\\
Log$_{10}$ water-rich planet pdf scale & $s_w$ & $\log_{10}$ \gcm & N(0.0, 0.35)\\
Log$_{10}$ sub-Neptune pdf scale & $s_p$ & $\log_{10}$ \gcm & N(0.0, 0.35)\\
\bottomrule
\end{tabular*}
\end{table}

The full parameterisation of the analytical model is shown in Table~\ref{table:model_parameters}. As mentioned, the rocky-planet transition start, $r_1$, stands for the radius below which all planets are rocky, while the sub-Neptune transition end, $r_4$,  stands for the radius above which all the planets are puffy Neptune-like planets. The water world population strength and shape parameters, $\pww$ and $\pwc$, define the water-rich planet population and are mapped to $r_2$ and $r_3$ as
\begin{align}
r_2 &= r_1 + (r_4 - r_1) (1 - \pww + \pwc\alpha), \label{eq:rtwo}\\
r_3 &= r_1 + (r_4 - r_1) (\pww + \pwc\alpha), \label{eq:rthree}
\end{align} 
where $\alpha = 0.5 - |\pww - 0.5|$. As shown in Fig.~\ref{fig:mixture_model}, the mapping is chosen so that the model can represent scenarios where water-rich planets do not form a separate composition class of their own. For $\pww = 0$, $r_1 = r_3$ and $r_2 = r_4$, and the rocky planet distribution transitions to the sub-Neptune distribution without a water world population in between; for $\pww = 0.5$, the water world population weight reaches unity at a single point between $r_1$ and $r_4$; and for $\pww = 1$, $r_1 = r_2$ and $r_3 = r_4$, and the water world population weight is unity for all radii between $r_1$ and $r_4$.\!\footnote{More details and examples about the analytical model weights and their parameterisation can be found from a Jupyter notebook at \url{https://github.com/hpparvi/spright/blob/main/notebooks/A1_analytical_model_weights.ipynb}.}

The rocky planet iron ratio and water-rich planet water ratio, $a$ and $b$, respectively, map directly to the iron and water mass fractions in the \citet{Aguichine21} and \citet{Zeng2019} models, and are used to interpolate the radius-density curves from the models.
The parameters $c$ and $d$ define the location and exponent of the power-law mean function of the sub-Neptune distribution, respectively. Finally, the three pdf scale parameters define the log-scales of the Student's t-distributions.

\subsection{Numerical radius-density probability model}
\label{sec:methods.numerical_model}
\subsubsection{Motivation}
\label{sec:methods.numerical_model.overview}

Figure \ref{fig:mixture_model} shows three individual realisations of the analytical mixture model that all agree with the observed radius-density distribution for small planets around M~dwarfs within the observational uncertainties. Instead of choosing to use the best-fit model, we opt for a more robust approach and calculate a numerical radius-density probability model that is constructed by averaging the analytical model over its posterior parameter space. That is, the numerical model corresponds to the mean posterior predictive distribution \citep{Gelman2013} of the analytical model given a set of planetary radius and mass observations with their uncertainties.

This approach has several advantages. On the one hand, it avoids both the rigidity arising from representing the M-R relationship as a simple parametric model and the need for large sample sizes necessary to make non-parametric models reliable. On the other hand, it lets the underlying analytical model take advantage of the theoretical radius-density models for rocky and water-rich planets by \citet{Zeng2019} and \citet{Aguichine21} but still ensures that the numerical model is not critically sensitive to the mean functions. The choice of these mean functions offers additional physical interpretability since the model can be parameterised by iron-rock and water-rock ratios, but they could be replaced by power laws --- or a different set of planetary internal structure models \citep[e.g.,][]{Dorn15, Mousis20} --- with relatively minor effects on the posterior model. Further, the approach allows us to also average over different mean density models to increase the robustness of our prediction.\!\footnote{Averaging the \package predictions over different theoretical mean density models can be carried out manually at the time of writing, but it will be added as an automated feature in the near future.}

\subsubsection{Likelihood}
\label{sec:methods.numerical_model.likelihood}

\begin{figure}
    \centering
    \includegraphics[width=\columnwidth]{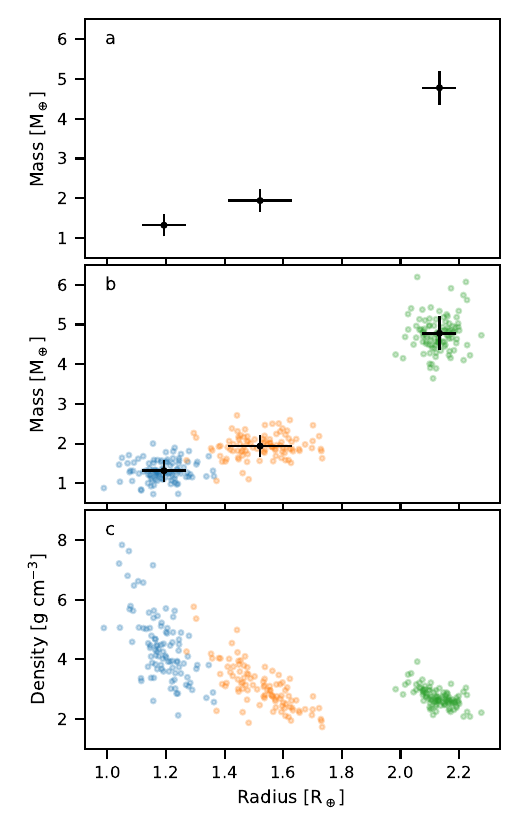}
    \caption{Generation of the mass and density samples for the likelihood evaluation. Panel a) depicts three planetary radius and mass measurements with their uncertainties, panel b) shows a set of radius and mass samples generated for each planet, and panel c) shows the radius and density samples derived from the radius and mass samples that are used in the model likelihood evaluation.}
    \label{fig:sample_creation}
\end{figure}

Ignoring observational uncertainties, our analytical radius-density probability model would give the log likelihood directly as a sum of the log probabilities of the $m$ observed ($r$, $\rho$) points as 
\begin{equation}
    \log L = \sum_{j=1}^m \log P(\rho_j | r_j, \vec{\theta}).
\end{equation}
However, the planet radius estimates from transit observations and mass estimates from RV observations have significant uncertainties that must be considered in the likelihood model. 
The code takes the uncertainties into account by drawing $n$ sets of radius and mass samples for each planet from the probability distributions defined by the observational uncertainties and transforming the mass samples into densities, as shown in Fig.~\ref{fig:sample_creation}. After this, the code adopts the likelihood averaged over the $n$ samples for each observation as the model log-likelihood,  
\begin{equation}
    \log L = \sum_{j=1}^m \log \frac{\sum_{i=1}^n P(\rho_\mathrm{i,j} | r_\mathrm{i,j}, \vec{\theta})}{n}. \label{eq:logl}
\end{equation}

For now, the mass and radius estimate uncertainties are assumed to be normal and symmetric, but we are planning to allow the observations to be represented by freely chosen probability distributions. This will allow, for example, the use of observations with only an upper limit on the planetary mass.

\subsubsection{Priors}
\label{sec:methods.numerical_model.priors}

We list the priors for the analytical model parameters in Table~\ref{table:model_parameters}.  The parameters controlling the transitions and the rocky planet iron-rock ratio have uninformative (uniform) priors. For \citet{Zeng2019}, the rocky planets with low Fe content are degenerate with water-rich planets with low H$_2$O content. To ensure that the water-world population actually represents water-rich planets,  we set a normal prior centred at 0.5 with a width of 0.1 for the H$_2$O-rock mixing ratio when using the \citet{Zeng2019} models. The meaning of the H$_2$O-rock mixing ratio is different between the \cite{Zeng2019} and \citet{Aguichine21} models (condensed versus supercritical water), and the latter do not have a problem with degeneracy with the \citet{Zeng2019} rocky-planet models. Because of this, we lift the normal prior constraint used with the \citet{Zeng2019} water-rich planet models and use a wide uniform prior on the H$_2$O-rock mixing ratio when using the \citet{Aguichine21} models. The sub-Neptune density normalisation factor and exponent have uninformative priors, and the logarithms of the Student's t-distribution widths' logarithms have loosely constraining normal priors. 

The parameters defining the transition regions of the model are given additional constraints to ensure that $r_1 \leq r_4$ and that the sub-Neptune density at the beginning of the sub-Neptune population ($r_3$) is never larger than rocky-planet density at the same radius.

\subsubsection{Creation of the numerical model}
\label{sec:methods.numerical_model.creation}

\begin{figure}
    \centering
    \includegraphics[width=\columnwidth]{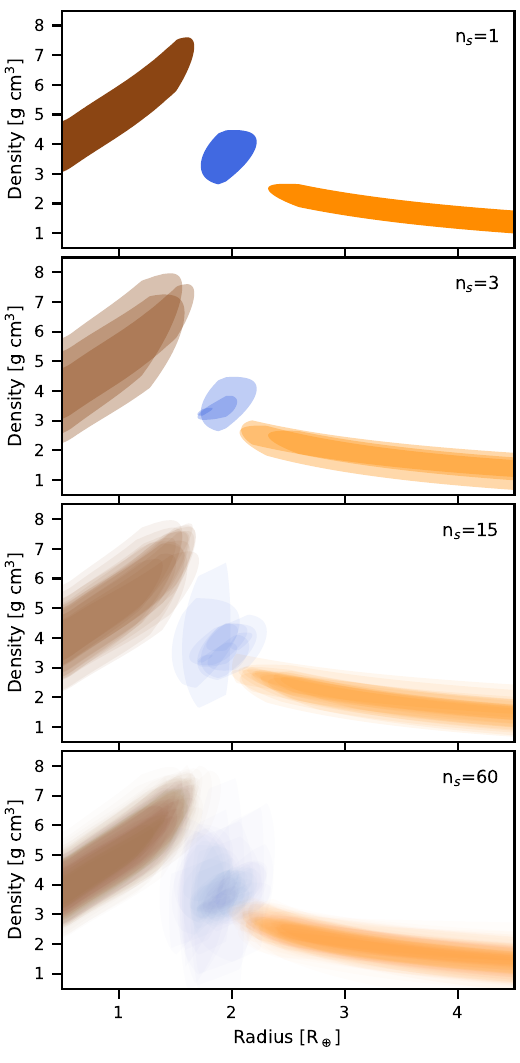}
    \caption{Construction process of the numerical radius-density probability model. The panels exhibit the mean values obtained from $n_\mathrm{s}$ samples of the analytical mixture model, drawn from its posterior distribution. In the visualisation, the rocky planet population is represented by brown colour, the water world population by blue, and the sub-Neptune population by orange. It is important to note that the figure presents the averages of a single isocontour for each component for visual clarity, whereas the actual model considers averages over real-valued probabilities. The default models included with the \package have been averaged over 3000 posterior samples.}
    \label{fig:mixture_model_averaging}
\end{figure}

When building the final numerical radius-density model, \package first finds the global  mode of the posterior density given the observational radius and mass estimates,
\begin{equation}
    \log P(\vec{\theta}|r,\rho) = \log L + \log P(\vec{\theta}),
\end{equation}
where $\log L$ is the log-likelihood from Eq.~\ref{eq:logl} and $\log  P(\vec{\theta})$ is the log-prior. The optimisation is carried out using the Differential Evolution global optimisation method \citep[DE, ][]{Price2005} implemented in \texttt{PyTransit} \citep{Parviainen2015}, with the initial parameter vector population drawn from the model parameter prior distribution. After the optimisation, the code obtains a sample from the model parameter posterior using the \emcee Markov chain Monte Carlo sampler \citep{Foreman-Mackey2012}. The \emcee sampler is initialised using the parameter population clumped around the global posterior mode by the DE method, and the sampler is run until it has obtained a representative sample from the posterior distribution.\!\footnote{The number of total and warm-up iterations are defined by the user when calculating a new model. The models included with the package used 60000 iterations in total with a warm-up period of 50000 iterations, where the quality of the final samples was confirmed by inspecting the evolution of the chain population.}
Next, the code discretises the radius-density space into a two-dimensional array and computes the posterior probability for each discrete ($r, \rho$) point by averaging the analytical probability model over the posterior samples,
\begin{equation}
    P(r, \rho) = \frac{1}{n} \sum_{i=1}^n P(r,\rho|\vec{\theta}_i),
\end{equation}
as illustrated in Fig.~\ref{fig:mixture_model_averaging}.

\begin{figure}
    \centering
    \includegraphics[width=\columnwidth]{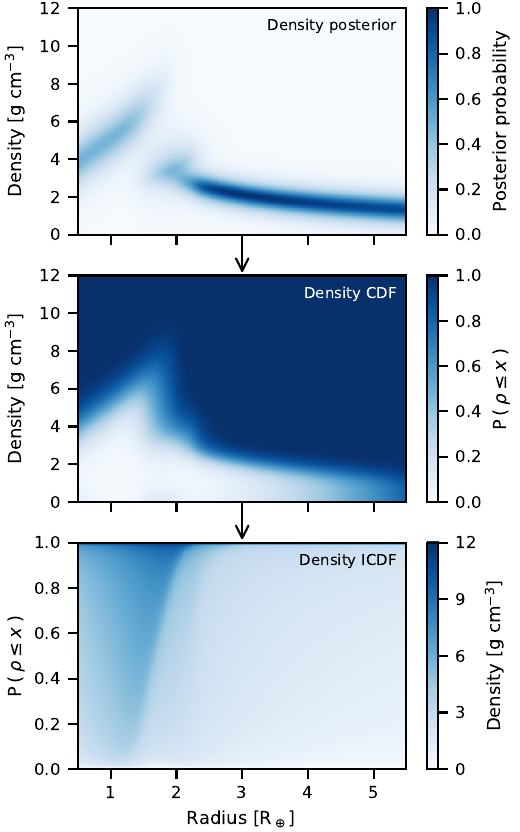}
    \caption{Generation of the inverse cumulative distribution function for the planetary bulk density as a function of the planet radius. The code starts with the numerical radius-density posterior model (top panel), calculates the cumulative distribution function for the bulk density as a function of planet radius (middle panel), and inverts it into an inverse cumulative distribution function (bottom panel).}
    \label{fig:posterior_to_density}
\end{figure}

After computing the numerical radius-density probability table, the code computes a discretised cumulative distribution function (CDF) for the planetary bulk density as a function of planet radius and, from this, a discretised inverse cumulative distribution function (ICDF) as a function of planet radius, as shown in Fig.~\ref{fig:posterior_to_density}. The posterior probability table, ICDF, parameter posterior samples, and the observational data are all then saved to a fits file used by the density and mass predicting part of the code. 

The creation of the numerical model is relatively fast, even for large radius and mass data sets, and scales linearly with the number of samples. The evaluation of the log-likelihood function (Eq.~\ref{eq:logl}) is parallelised to take advantage of modern multi-core processors, so that, for example, calculating a new model for a data set with 158 planets (i.e. the TEPCat FGK catalogue discussed later) takes 3-7 minutes on a relatively modern eight-core desktop computer.

\subsubsection{Evaluation of the numerical model}
\label{sec:methods.numerical_model.evaluation}

\begin{figure}
    \centering
    \includegraphics[width=\columnwidth]{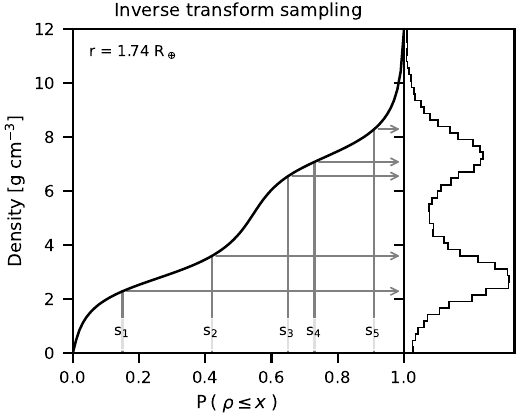}
    \caption{Inverse transform sampling in one dimension for a single value of planet radius, $r$. The inverse cumulative distribution function (ICDF) transforms a uniform distribution from 0 to 1 to the estimated bulk density distribution. The figure shows five samples from a uniform distribution (s$_1$ to s$_5$) and how they map to samples from the planet density distribution.}
    \label{fig:inverse_sampling_1d}
\end{figure}

\package uses inverse transform sampling (Fig.~\ref{fig:inverse_sampling_1d}) to draw a planet density sample given a planet radius with its uncertainties and a saved numerical radius-density model. The code draws a number of $(r,p)$ samples where $r$ follows from the planet radius distribution and $p \sim U(0,1)$ and obtains a density sample for each $(r,p)$ value by linearly interpolating the 2D ICDF table. After this, a mass sample is obtained from the density sample by multiplying the densities with the respective planet volumes.

The model evaluation is extremely fast since it consists only of the generation of $n$ samples from the two distributions followed by bilinear interpolation inside the ICDF table.

\begin{figure}
    \centering
    \includegraphics[width=\columnwidth]{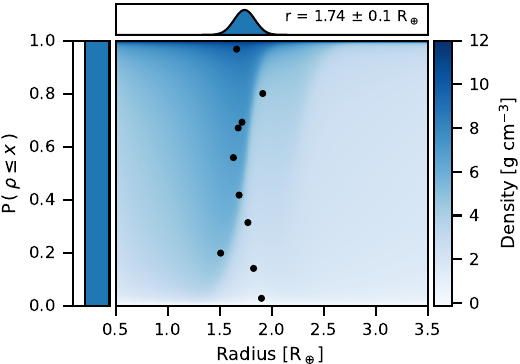}
    \caption{Inverse transform sampling in two dimensions where the planet radius contains uncertainty. The density samples are obtained by drawing samples from the planet radius distribution and a uniform distribution between 0 and 1 and evaluating the ICDF for each radius sample as in Fig.~\ref{fig:inverse_sampling_1d}. Because the ICDF is stored as a discrete two-dimensional array, the density sampling corresponds simply to bilinear interpolation inside the ICDF array.}
    \label{fig:inverse_sampling_2d}
\end{figure}

\subsection{Model usage}

\subsubsection{Model creation}
The creation of a new radius-density-mass relationship is carried out with the \texttt{spright.RMEstimator} class. At its simplest, the class can be initialised with the system names, a list of planetary radii with their uncertainties, and a list of planetary masses with their uncertainties. The initialisation is followed by model optimisation, parameter posterior estimation, and ICDF map computation: 
\begin{lstlisting}[language=Python]
from spright import RMEstimator

rme = RMEstimator(names=names, 
                  radii=radii, 
                  masses=masses)
rme.optimize()
rme.sample()
rme.compute_maps()
rme.save('map_name.fits')
\end{lstlisting}
After the ICDF map is computed, it can be saved to be used in model evaluation. The class also allows the model creation to be tuned for specific interests by changing parameter priors or setting additional constraints.

\subsubsection{Model evaluation}
\label{sec:methods.usage.evaluation}

\begin{figure}
    \centering
    \includegraphics[width=\columnwidth]{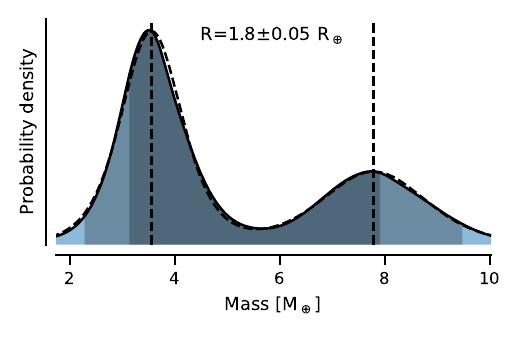}
    \caption{Predicted mass distribution for a planet with a radius of $1.8\pm0.05$ based on the updated STPM catalogue. The solid line shows the actual distribution, the dashed line shows a two-component mixture model fitted to the distribution, the dashed vertical lines show the mixture model component centres, and the grey shading shows three central posterior percentile limits for the distribution.}
    \label{fig:mass_distribution}
\end{figure}

After a radius-density model has been computed, it can be evaluated using the \texttt{spright.RMRelation} class. The class offers methods to predict the planet's bulk density, mass, and RV semi-amplitude\footnote{The prediction of RV semi-amplitude also requires an estimate for the stellar mass, orbital period, inclination, eccentricity, and argument of periastron.} distributions given its radius with uncertainties, or the planet's radius given its mass and its uncertainty. For example, a mass distribution for an $r=1.8\pm0.05 \rearth$ planet can be obtained as
\begin{lstlisting}[language=Python]
from spright import RMRelation

rmr = RMRelation('map_name.fits')
mds = rmr.sample('mass', (1.8, 0.05))
mds.plot()
\end{lstlisting}
where \texttt{mds} is a \texttt{spright.Distribution} object that offers utility methods for plotting the distribution, approximating it with analytical (mixture) distributions, calculating distribution percentiles, and so on. Figure~\ref{fig:mass_distribution} shows the plot created by the \texttt{mds.plot()} method visualising the actual numerical mass distribution, an analytical distribution fitted to the numerical distribution, and a set of distribution percentile limits. 

\subsubsection{Model evaluation from the command line}
\label{sec:methods.usage.evaluation_cl}

The package includes a command line tool \texttt{spright} that makes the model evaluation easy without any coding required. The example above can be evaluated from the command line as
\begin{lstlisting}[language=bash]
spright --predict mass --radius 1.8 0.05
\end{lstlisting}
where the script can also save a publication-quality plot of the predicted distribution and the radius-density map used to create the prediction.

\section{Discussion}

\subsection{Comparison between different catalogues}
\label{sec:discussion.catalogue_comparison}

\begin{figure*}
    \centering
    \includegraphics[width=\textwidth]{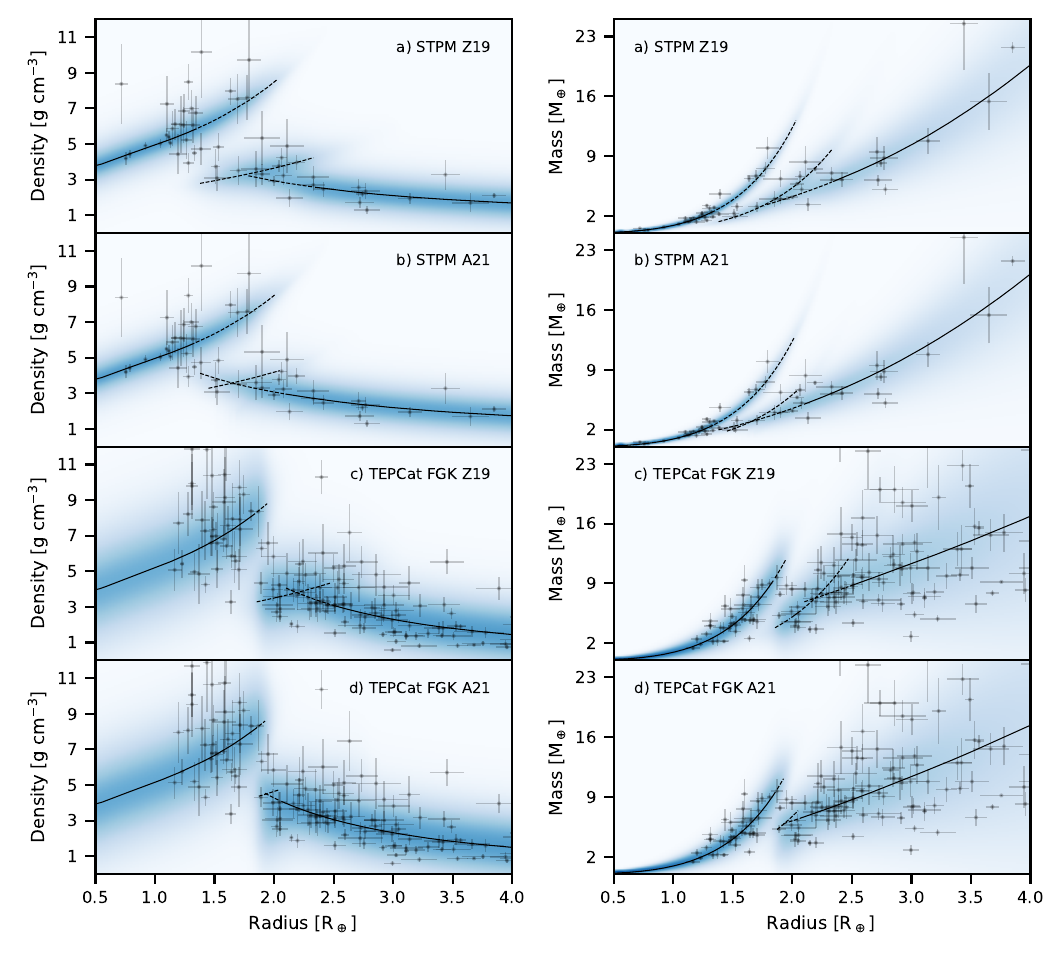}
    \caption{Numerical radius-density (left) and radius-mass (right) probability models fitted to the STPM M~dwarf catalogue and the TEPCat FKG star catalogues using either \citet[Z19, ]{Zeng2019} or \citet[A21, ][]{Aguichine21} water-rich planet density models to represent the density mean function for the water worlds, as explained in Sect.~\ref{sec:discussion.catalogue_comparison}. Grey data points show radius, density, and mass measurements with their uncertainties for all planets in each catalogue. The blue colour corresponds to the logarithm of the posterior probability, and the black lines show the posterior means for each of the three planet populations: the solid lines correspond to radius regimes where the component has a weight of unity (that is, all planets in this range belong to this component), while the dashed lines mark the transition regimes between the populations.}
    \label{fig:three_models}
\end{figure*}

The flexibility of \package allows the user to quickly generate a new joint radius-density probability distribution based on different empirical data sets. In this section, we use \package to compare the distributions obtained using two  catalogues of exoplanet properties and two sets of theoretical density models for the water-world population. The water-rich planet density models are the previously-mentioned models by \citet{Aguichine21} and \citet{Zeng2019}, and the two catalogues are: 
\begin{itemize}
    \item[a)] an updated\footnote{Including planets with measured radii and masses published between June 2021 and 2023. Available in the \package GitHub repository.} version of the Small Transiting Planets around M~dwarfs (STPM) catalogue by \citet{Luque2022} (containing 48 planets spanning a range of 168--1089\,K in equilibrium temperature and 0.089 to 0.63\,$M_\odot$ in stellar host mass),
    \item[b)] a sample of small planets orbiting FGK stars from the TEPCat catalogue \citep{tepcat} with better than 25\% and 8\% uncertainties in the mass and radius, respectively (containing 159 planets spanning a range of 380--2350\,K in equilibrium temperature and 0.59 to 1.26\,$M_\odot$ in stellar host mass).\!\footnote{The \package package also includes a model calculated for a sample of small planets ($R < 4\,\rearth$) around M dwarfs ($T_{\rm eff} < 4000\,\mathrm{K}$) taken from the TEPCat catalogue  (containing 54 planets spanning a range of 168--1323\,K in equilibrium temperature and 0.089 to 0.65\,$M_\odot$ in stellar host mass), but the inferred parameter posteriors are so similar to the ones inferred from the STPM catalogue that we do not include a comparison here.}
\end{itemize} 

Figure~\ref{fig:three_models} shows the radius-density and radius-mass distributions inferred from the two catalogues and water-world density models, and a comparison of the model parameter posterior distributions can be found in Figs.~\ref{fig:parameter_posteriors}, \ref{fig:parameter_joint_posteriors_stpm}, \ref{fig:parameter_joint_posteriors_fgk}, \ref{fig:derived_parameter_joint_posteriors_fgk}, \ref{fig:derived_parameter_joint_posteriors_stpm}, \ref{fig:parameter_joint_posteriors_stpm_a21}, \ref{fig:parameter_joint_posteriors_fgk_a21}, \ref{fig:derived_parameter_joint_posteriors_fgk_a21}, and \ref{fig:derived_parameter_joint_posteriors_stpm_a21}. The values of most of the model parameters are consistent regardless of the catalogue used. However, the uncertainties of the model parameters decrease for the TEPCat FGK catalogue (particularly those related to the water world and sub-Neptune transition), which is the one with the largest number of planets. This result highlights the importance of increasing not only the precision and accuracy of exoplanet observed properties but also the number of planets with those properties measured in such a manner. 

On the one hand, the iron-rock mixing ratios and the parameters defining the sub-Neptune population agree between the two water-rich planet density models, but differ between the STPM and TEPCat FGK catalogues. On the other hand, the water-rock mixing ratios differ between the density models but agree between the catalogues. The discrepancy in the water-rock mixing ratio between the A21 and Z19 models is however to be expected because the parameter has a different physical meaning for each model \citep[see][for details]{Zeng2019,Mousis20,Aguichine21,DornLichtenberg2021}.

The water world population seems to shift to higher radii for the FGK hosts compared to the M dwarfs. Type I migration models, independent of the solid accretion mechanism (planetesimal- or pebble-based), predict a planet-to-disk-mass dependent mass-scale where planets migrate \citep{Burn21, Schlecker22}. Thus, in the M-dwarf case, lower-mass water worlds are able to migrate rapidly enough to reach the distances probed in the catalogues compared to the solar-type hosts. This dichotomy in migration timescales could be responsible for the shift to larger sizes of the water world population observed in the FGK versus M dwarf sample.

Compared to the STPM, the TEPCat FGK catalogue shows a stronger separate water-rich population from rocky worlds. The transition is also sharper, indicating a smaller overlap between rocky and water worlds. As discussed above, this effect can also be understood as a consequence of the higher minimum-mass water-rich planets able to migrate to the inner parts of the system for FGK hosts compared with M dwarfs. The lack of low-mass water-rich planets in the FGK sample, however, can be related to an observational bias. Water-rich planets around FGK stars with radii below $2\,\rearth$ generally have RV semi-amplitudes of 1~m/s or smaller,\!\footnote{Based on the \citet{Aguichine21} and \citet{Zeng2019} water-rich planet density models.} which limits their mass estimation with current RV instrumentation. Observational and detectability biases manifest also in the FGK sample at the Earth- and sub-Earth-size limit, where the lack of rocky planets with measured bulk densities is not because they are intrinsically rare \citep[as clearly demonstrated by, e.g.,][]{Batalha13}, but due to their hardly detectable sub-meter-per-second RV signals.

\subsection{Comparison with previous mass-radius relations}

\begin{figure*}
    \centering
    \includegraphics[width=\textwidth]{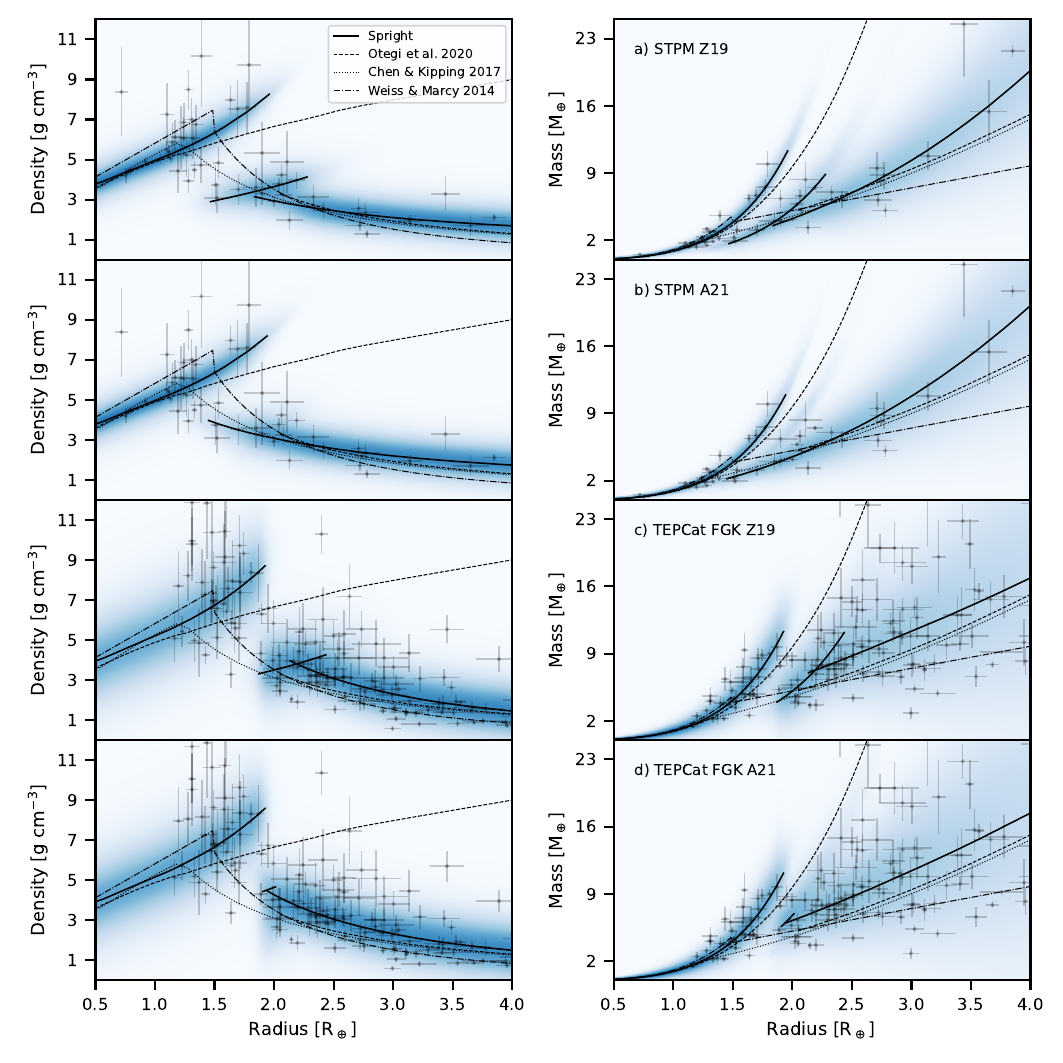}
    \caption{The \package radius-density and radius-mass relations for the STPM and TEPCat catalogues and \citet[A21, ][]{Aguichine21} and \citet[Z19, ][]{Zeng2019} models for the water-rich planet densities plotted together with the ones by \citet{Otegi20}, \citet{ChenKipping17}, and \citet{Weiss2014}. The solid lines correspond to the \package posterior distribution means for planet radii where the component weight is larger than 0.1, the slashed lines to the \citet{Otegi20} means, the dotted lines to the \citet{ChenKipping17} means, and the dash-dotted lines to the \citet{Weiss2014} relation. The blue shading shows the \package joint probability densities. We do not show the model uncertainties for visual clarity.}
    \label{fig:mr_comparisons}
\end{figure*}

A significant amount of work has been invested during the last years to model the relationship between planetary radii and masses \citep{Lissauer11, Weiss2014, Wolfgang2016, Mills2017, ChenKipping17, Bashi17, Ning2018, Kanodia2019, Otegi20}. Most of these works have followed a parametric approach, modelling the mass-radius dependence as one or multiple power-law segments \citep[e.g.,][]{Weiss2014, Wolfgang2016, ChenKipping17, Otegi20}; although non-parametric approaches have also been explored \citep[e.g.,][]{Ning2018, Kanodia2019}. The numerical M-R relation offered by \package aims to overcome the shortcomings of the existing parametric and non-parametric models. It offers the flexibility and robustness of non-parametric approaches (in particular, modelling the \emph{joint} radius-density distribution rather than a single variable) while being conceptually and computationally simple to interpret and incorporate into a Bayesian framework. Figure~\ref{fig:mr_comparisons} shows the radius-density and radius-mass relations obtained with \package compared to others that are widely used in the community \citep[e.g.,][]{Weiss2014, ChenKipping17, Otegi20}. 

Among these works, the results obtained with \package are particularly consistent with those by \citet{Otegi20}. \citet{Otegi20} introduced two separate power law components based on the division in mass-radius space set by the equation of state of water \citep{Dorn15}, allowing a better representation of the transition region between rocky and volatile-rich planets. The agreement between their volatile-rich power law segment and our hydrogen-rich model is remarkable. But, their rocky component seems to overestimate the size of the rocky planets with masses between 5 to $10\,M_\oplus$. Furthermore, contrary to \package, the M-R relations from \citet{Otegi20} require prior knowledge of the planet's bulk density to choose the adequate power-law, which is key to adequately predict the mass of the planets with radii between 1.5 and $3\,R_\oplus$.

Regarding the other models, \package obtains consistent results for the rocky population compared to \citet{Weiss2014} and \citet{ChenKipping17} up to approximately $1.5\,\rearth$. However, \citet{ChenKipping17} sets the transition between rocky and volatile-rich planets at $2\,M_\oplus$, thus failing to reproduce the high-mass tail of the rocky population with masses between 2 and $10\,M_\oplus$ that overlaps with the water worlds and is particularly prominent in the FGK catalogue. This difference is partially due to our use of new mass and radius data that was not available when the \texttt{Forecaster} model was fit, and partially due to \texttt{Forecaster} using a piecewise model to represent the mass-radius relation, while \package is using a mixture model. That is, \texttt{Forecaster} cannot model a situation with two overlapping populations, but \package assumes by default that the populations can overlap. Outside of the low-mass range, the agreement is very good. In particular, for the hydrogen-rich sub-Neptune population, \package infers a power law index of $-0.8\pm0.3$ while \citet{ChenKipping17} obtained $-0.77\pm0.13$. Conversely, \citet{Weiss2014} fails to predict the properties of the volatile-rich population (overestimating the average size and, consequently, underestimating the density of this type of planet). 

\subsection{Evidence for water worlds}
\label{sec:discussion.water_worlds}

The analytical mixture model can explain the observed radius and mass distribution either as a mixture of rocky planets and sub-Neptunes, or as a mixture of rocky planets, water-rich planets, and sub-Neptunes (see Fig.~\ref{fig:mixture_model}). This flexibility allows the model to be used for studying whether observations support or contradict the presence of water-rich worlds as a distinct planetary population positioned between rocky planets and sub-Neptunes.

The analytical model uses the water world population strength, $\pww$, to parameterise the significance of the water world population: $\pww = 0$ corresponds to a model without water worlds, $\pww = 0.5$ corresponds to a model where the water world population weight reaches unity for a single point in the planet radius space, and $0.5 < \pww \leq 1$ correspond to models where the water world population weight is unity for a fraction of the transition region between rocky planets and sub-Neptunes. This parametrisation allows us to present three competing hypotheses:
\begin{description}
    \item[H$_0$)] water worlds do not exist as a distinct population,
    \item[H$_1$)] water worlds exist as a mixed population, or
    \item[H$_2$)] water worlds exist a significant distinct population, 
\end{description}
merely by setting different priors on \pww. For H$_0$, we can force the model to exclude the water world population by setting a delta function prior, $\delta(0)$ on \pww. The mixed-population hypothesis, H$_1$, represents a scenario where a population of water worlds exists mixed with rocky planets and sub-Neptunes, but there are no planetary radii for which all the planets would be purely water worlds. We encode H$_1$ by a uniform prior on \pww from 0.015 to 0.5, where the lower bound represents a weak water world population and the upper bound represents the case where the water world population weight reaches unity for a single point in the planet radius space. Finally, for H$_2$, we choose to encode a "significant distinct population" by a uniform prior from 0.5 to 1.0. That is, the water world population weight must reach unity for at least one point in the radius space, and the upper limit marks the model with sharp transitions from rocky planets to water worlds and from water worlds to sub-Neptunes.

After defining our hypotheses, H$_0$, H$_1$ and H$_2$, we can calculate their respective Bayesian evidences, Z$_0$, Z$_1$ and Z$_2$, by integrating over the model posterior spaces \citep{Parviainen2018, Gelman2013, Robert2007, Kass1995}.\!\footnote{The \texttt{Jupyter} notebooks used to estimate the Bayesian evidences are available from the \package GitHub repository. We implement the $\delta$ prior on \pww for H$_0$ as a narrow uniform prior from 0 to 0.001.} We carry out the integration using the \texttt{Dynesty} package \citep{Koposov2023, Speagle2020, Feroz2009, Skilling2006, Skilling2004} to estimate the Bayesian evidences via dynamical nested sampling. 

\begin{table}
    \centering
    \begin{tabularx}{\columnwidth}{Xr@{ $\pm$ }lr@{ $\pm$ }lr@{ $\pm$ }l}
    \toprule\toprule
    Catalogue & \multicolumn{2}{c}{$\bfactor_{10}$}  & \multicolumn{2}{c}{$\bfactor_{20}$} & \multicolumn{2}{c}{$\bfactor_{21}$} \\
    \midrule
        STPM Z19& -0.6 & 0.7 & -3.4 &  0.7 & -2.8 & 0.7 \\
        STPM A21& -7.5 & 0.6 & -8.5 &  0.7 & -1.0 & 0.6 \\
        TEPCat FGK Z19 & 2.5 &  0.7 &  2.6 &  0.7 & 0.2 & 0.7\\
        TEPCat FGK A21 &-1.0 &  0.7 & -1.5 &  0.7 & -0.5 & 0.7 \\
    \bottomrule
    \end{tabularx}
    \caption{Bayes factors inferred from the STPM and TEPCat FGK catalogues for two theoretical water-rich-planet mean radius-density models and three hypotheses detailed in Sect.~\ref{sec:discussion.water_worlds}. The Z19 scenarios use the \citet{Zeng2019} models to represent the mean radius-density function of the water-rich-planet population, while the A21 scenarios use the models by \citet{Aguichine21}.}
    \label{table:evidence}
\end{table}

We report the log Bayes factors --- defined as $2\ln B_{10} = 2\ln Z_1 - 2\ln Z_0$ and $2\ln B_{20} = 2\ln Z_2 - 2\ln Z_0$ --- in Table~\ref{table:evidence}. Assuming the \citet{Zeng2019} water-rich planet density models, the evidence is insufficient to significantly support any of the scenarios over the others \citep[][p.~777]{Kass1995}.\!\footnote{We adopt the Bayes evidence interpretation of \citet{Kass1995}, where $\bfactor_{10} < 2$ is considered as insignificant, $2 \leq \bfactor_{10} < 6$ as positive, $6 \leq \bfactor_{10} < 10$ as strong, and $\bfactor_{10} \geq 10$ as very strong evidence against H$_0$.} Assuming the \citet{Aguichine21} models, H$_0$ (no water world population) is strongly favoured over H$_1$ and H$_2$ for the STPM catalogue, but only tentatively favoured for the TEPCat FGK catalogue. The reason for this discrepancy between Z19 and A21 models for the STPM catalogue comes from the impossibility of A21 models to match the slope of the water-rich planet population in mass-radius space for low-mass planets ($< 5\,M_\oplus$), while it is easily reproduced by Z19 models. On the other hand, the higher average planet radius of the FGK sample is easier to reproduce with the A21 models compared to Z19. We note that \package only uses a set of A21 models that assume $T_{\rm irr}=500$\,K and an Earth-like composition for the core. A thorough exploration of our model comparison results as a function of these two parameters is beyond the scope of this paper.

\subsection{Synthetic catalogue tests}
\subsubsection{Parameter posteriors}
\label{sec:discussion.mock.posteriors}
While the main use case for \package is in predicting planet masses given their radii and vice versa, the model parameter posteriors can also be used to study the physical properties of small-planet populations. For this use, however, we need to understand how the posteriors depend on factors such as the number of planets included in the model calculation.

To this end, we carried out a synthetic catalogue study for all combinations of four catalogue sizes (the number of planets included in the catalogue, $\catsize \in \{50, 100, 150, 200\}$) and three values for the water world population strength ($\pww \in \{0.0, 0.5, 1.0\}$). We created five realisations of synthetic (radius, mass) catalogues with realistic uncertainties on both quantities for each (\catsize,  \pww) combination. The parameters other than \pww were fixed to $r_1=1.3$, $r_4=2.4$, $\phi=0.0$, $a=0.2$, $b=0.5$, $c=3$, $d=-0.9$, $s_\mathrm{r}=-0.4$, $s_\mathrm{w}=-0.3$, and $s_\mathrm{p}=-0.4$. The synthetic mass and radius data sets were created using the \texttt{create\_mock\_sample} helper function in \package, and the relative radius uncertainties were drawn from a uniform distribution from 1\% and 8\%, while the relative mass uncertainties were drawn from a uniform distribution from 3\% and 24\%. To simplify the analysis, we only use the \citet{Zeng2019} density models for the water-rich planets.

\begin{figure*}
    \centering
    \includegraphics[width=\linewidth]{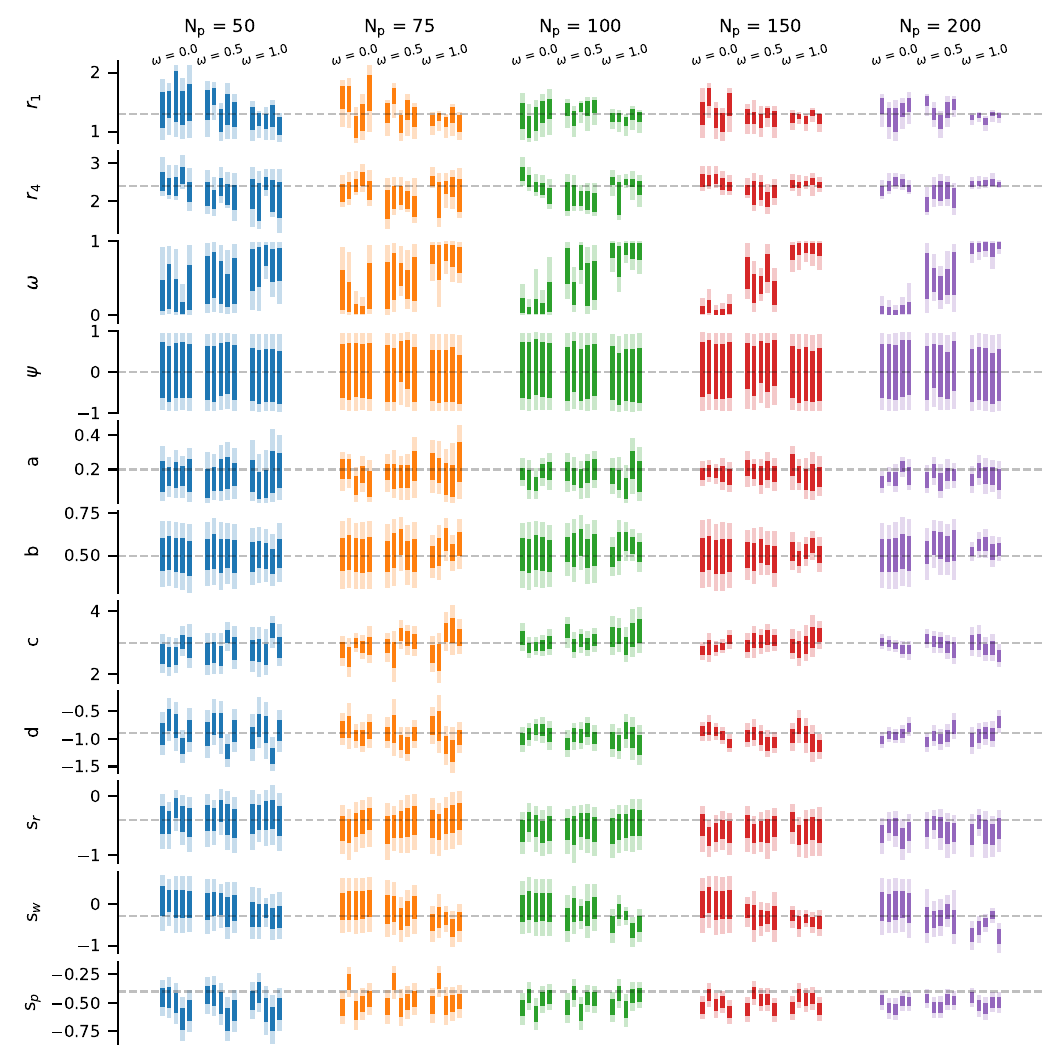}
    \caption{Posterior distributions for all the \package model parameters inferred from the 75 synthetic mass and radius catalogues described in Sect.~\ref{sec:discussion.mock.posteriors}. The light and dark vertical bars show the central 95\% and 68\% posterior limits, respectively, and the horizontal dotted line shows the true value for all parameters except the water world population strength, \pww. The simulations were carried out for five catalogue sizes ($N_\mathrm{p}$) and three values of \pww, and the posteriors are grouped first by $N_\mathrm{p}$ (outer level grouping with separate colour for each $N_\mathrm{p}$), then by \pww (a set of five posterior estimates), and finally by the catalogue realisation (a single vertical line).}
    \label{fig:mock_data_parameter_posteriors}
\end{figure*}

We created the numerical \package model for each synthetic catalogue, and show the inferred parameter posteriors in Fig.~\ref{fig:mock_data_parameter_posteriors}. The true parameter values are generally contained inside the inferred 68\% central posterior limits, and only very rarely outside the 95\% central posterior limits.  The posterior uncertainties decrease with the increasing catalogue size as expected, except for the water-world population shape parameter, \pwc. This is not entirely surprising because \pww has the strongest effect on the shape of the water-world population for intermediate values of \pww and it will likely be well-constrained only for relatively strong water-world populations and large catalogue sizes.

\begin{figure}
    \centering
    \includegraphics[width=\linewidth]{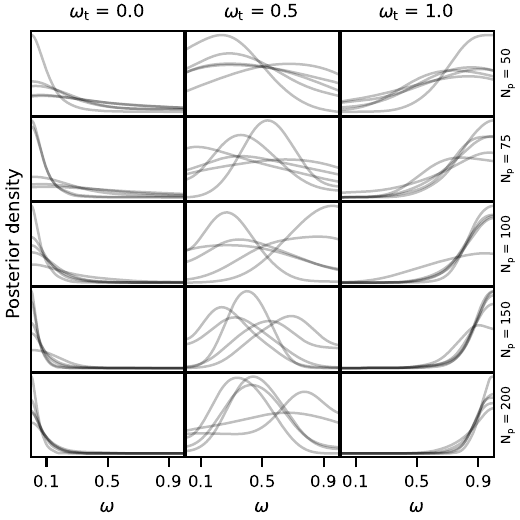}
    \caption{Posterior distributions for the water world population strength \pww inferred from the 75 synthetic mass and radius catalogues described in Sect.~\ref{sec:discussion.mock.posteriors}. The columns show the posteriors for a given $\pww_\mathrm{t}$ (where the subscript stands for "truth"), and the rows for a different catalogue size,  $N_\mathrm{p}$.}
    \label{fig:mock_ww_posteriors}
\end{figure}

We also studied how the posterior estimate for \pww changes for the three simulated \pww scenarios and five catalogue sizes, and show the results in Fig.~\ref{fig:mock_ww_posteriors}. The water world population strength is relatively well constrained for both extreme cases even for $N_\mathrm{p} = 50$, and the distribution mode is in all cases close to the true \pww value. The intermediate $\pww = 0.5$ scenario is less well constrained, but the posteriors generally differ from the posteriors from the extreme cases. In practice, this means that the model can distinguish between the two extreme cases for \pww, and a poorly-constrained \pww can be interpreted as support for the existence of a population of water-rich planets that is mixed with the rocky and sub-Neptune populations.

\subsubsection{Water-world evidence}
\label{sec:discussion.mock.evidence}

We repeated the Bayesian model comparison test in Sect.~\ref{sec:discussion.water_worlds} using synthetic catalogues. We calculated the Bayesian evidences for the H$_0$, H$_1$, and H$_2$ hypotheses for 20 catalogue realisations for each (\catsize, \pww) combination, where $\catsize \in \{50, 100, 150, 200\}$ and $\pww \in \{0.0, 0.25, 0.5, 0.75, 1.0\}$, again restricting the simulations to the \citet{Zeng2019} density model for the water-rich planets.\!\footnote{We will repeat the simulation for the \citet{Aguichine21} water-rich planet density models in the future and make the results publicly available from the \package GitHub repository, but this work is beyond the scope of this paper.}

\begin{figure}
    \centering
    \includegraphics[width=\linewidth]{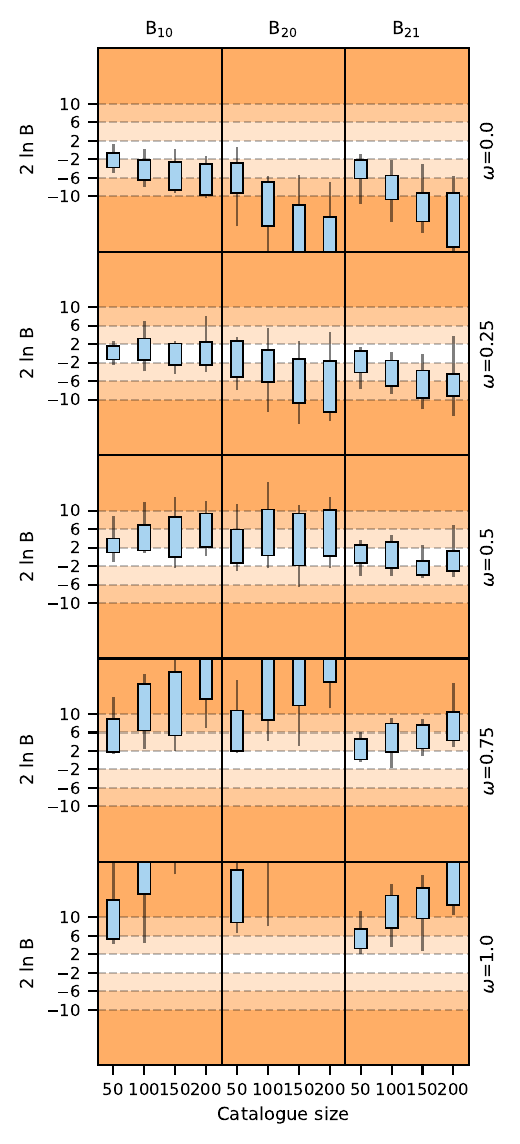}
    \caption{Log Bayes factors from the synthetic catalogue study described in Sect~\ref{sec:discussion.mock.evidence}. The columns show the Bayes factors for hypotheses H$_1$ and H$_2$ in favour of H$_0$, and H$_2$ in favour of H$_1$, and the rows show the different \pww scenarios. The number of planets in the catalogue is shown on the x-axis, and the y-axis shows the 68\% central intervals of the \bfactor value distributions as blue boxes while  black vertical lines mark the distribution minimum-to-maximum spans. The yellow shading follows the four levels of support by \citet{Kass1995}: $2\ln B < 2$) insignificant support, $2 < 2\ln B < 6$) positive support, $6 < 2\ln B < 10$) strong support, and $2\ln B > 10$) very strong support. Negative $\ln B$ values mean the alternative hypothesis is favoured, so that $2\ln B_{10} < -10$ would mean very strong support for H$_0$ over H$_1$.}
    \label{fig:mock_bayes_factors}
\end{figure}

We visualise the resulting evidence distributions in Fig.~\ref{fig:mock_bayes_factors} and summarise them in Table~\ref{table:mock_evidence}.  
For $\pww=0$, the log Bayes factors, $\bfactor_{10}$ and $\bfactor_{20}$, generally support the no-water-world-population hypothesis H$_0$ over the two others. The evidence against H$_1$ is rather tentative for small \catsize, and can be weak even for large \catsize. The strong-water-world-population scenario, H$_2$, can be ruled out in most cases with "positive" evidence already with small \catsize, and decisively with large \catsize. 
For $\pww=0.25$, the evidence for the mixed-water-world-population hypothesis, H$_1$, reaches the level of "strong" evidence for some catalogue realisations, but in most cases, neither H$_0$ nor H$_1$ is strongly favoured over another. Both H$_0$ and H$_1$ are generally favoured over H$_2$, but the level of support for H$_1$ does not increase significantly with increasing \catsize.
For $\pww=0.5$, $\bfactor_{10}$ is nearly always positive and reaches high levels of evidence for larger \catsize, while the support for H$_2$ over H$_0$ varies from slightly negative to strongly positive.  
For $\pww=0.75$, both H$_1$ and H$_2$ are significantly favoured over H$_0$ for all \catsize, and H$_2$ is generally favoured over H$_1$. 
For $\pww=1$, The strong-water-world-population hypothesis H$_2$ is favoured over H$_0$ and H$_1$ with decisive support already for $\catsize=50$.

All in all, the synthetic tests show that the \package model can be used to distinguish between the two extreme cases described by \pww values of 0 and 1 already with a catalogue consisting of $\sim 50$ planets. Interestingly, the scenario with $\pww=1$ can be identified much more securely than the scenario with $\pww=0$. In most cases, $\bfactor_{20} < -2$ for $\pww=0$ and  $\bfactor_{20} > 6$ for $\pww=1$, and the contrast between H$_0$ and H$_2$ increases quickly together with the number of planets included into the catalogue. The intermediate cases with \pww values of 0.25 and 0.5 are identified less securely, but for $\pww\sim0.5$, $\bfactor_{10}$ is still generally positive while being negative for $\pww=0$. 

\begin{table}
    \centering
\begin{tabular*}{\columnwidth}{@{\extracolsep{\fill}} llrrrr}
\toprule\toprule
$N_\mathrm{p}$ & $\omega$ & min $\bfactor_{10}$  & max $\bfactor_{10}$  & min $\bfactor_{20}$ & max $\bfactor_{20}$ \\
\multirow[t]{5}{*}{50} & 0.0 & -4.8 & 1.4 & -16.5 & 0.6 \\
 & 0.25 & -2.4 & 2.8 & -7.9 & 3.5 \\
 & 0.50 & -1.2 & 8.8 & -3.1 & 11.4 \\
 & 0.75 & 1.4 & 13.7 & 1.6 & 17.3 \\
 & 1.00 & 4.3 & 22.3 & 6.7 & 33.7 \\
\multirow[t]{5}{*}{100} & 0.0 & -8.0 & 0.1 & -23.5 & -5.5 \\
 & 0.25 & -3.8 & 7.0 & -12.6 & 5.5 \\
 & 0.50 & 0.9 & 11.9 & -2.5 & 16.3 \\
 & 0.75 & 2.5 & 18.6 & 4.2 & 26.8 \\
 & 1.00 & 4.4 & 32.0 & 8.0 & 48.7 \\
\multirow[t]{5}{*}{150} & 0.0 & -9.2 & 0.1 & -26.6 & -5.3 \\
 & 0.25 & -4.4 & 2.7 & -15.2 & 2.6 \\
 & 0.50 & -2.4 & 13.0 & -6.5 & 11.3 \\
 & 0.75 & 1.9 & 27.8 & 3.2 & 36.8 \\
 & 1.00 & 19.3 & 46.2 & 23.9 & 62.9 \\
\multirow[t]{5}{*}{200} & 0.0 & -10.3 & -1.3 & -31.6 & -6.9 \\
 & 0.25 & -4.0 & 8.1 & -14.7 & 4.6 \\
 & 0.50 & 0.1 & 12.0 & -2.3 & 13.0 \\
 & 0.75 & 7.0 & 30.8 & 11.2 & 41.3 \\
 & 1.00 & 26.6 & 67.2 & 38.7 & 91.5 \\
\bottomrule
\end{tabular*}
\caption{Minimum and maximum Bayes factors (\bfactor) for hypotheses H1 and H2 in favour of H0 from the synthetic catalogue simulations visualised in Fig.~\ref{fig:mock_bayes_factors}.}
\label{table:mock_evidence}
\end{table}

\section{Conclusions}

We have presented \package,\!\footnote{\url{https://github.com/hpparvi/spright}, \href{https://doi.org/10.5281/zenodo.10082653}{DOI:10.5281/zenodo.10082653 }} a \texttt{Python} package that implements a lightweight probabilistic radius-density-mass relationship for small planets based on basic Bayesian inference. The package represents the joint planetary radius and bulk density distribution as a mean of the posterior predictive distribution of a simple analytical three-component mixture model. The package offers tools to predict planetary masses, bulk densities, and radial velocity semi-amplitudes for planets orbiting M~dwarfs (based on the revised STPM catalogue by \citealt{Luque2022}) and FGK~stars (based on TEPCat catalogue). The package has been designed to be easy to install and use and also aims to make the computation of new M-R relations easy. Further, calculating a new M-R model takes only minutes with a modern desktop computer, even for large data sets containing hundreds of planets, and the computing time scales linearly with the number of planets.

We have also studied whether the current observational radius and mass estimates support the existence of water-rich worlds as a distinct planet population between rocky planets and sub-Neptunes. While the numerical M-R model is agnostic to what comes to the existence of water worlds, the analytical model can be used in a Bayesian model comparison setting to assess the level of evidence in favour of a distinct water world population. Our study finds that the inferred support for the existence of a water world population depends on the chosen theoretical water-rich planet density model. All in all, the TEPCat data set is insufficient to provide statistically significant evidence for or against the existence of a water world population around FGK stars. The STPM data set shows some evidence against the existence of a water world population around M~dwarfs, but this evidence is not strong enough to be considered conclusive, in line with the recent results from \citet{Rogers2023}. 

\section*{Acknowledgements}
HP acknowledges support from the Spanish Ministry of Science and Innovation with the Ramon y Cajal fellowship number RYC2021-031798-I, as well as funding from the University of La Laguna and the Spanish Ministry of Universities. RL acknowledges funding from the University of La Laguna through the Margarita Salas Fellowship from the Spanish Ministry of Universities ref. UNI/551/2021-May 26, and under the EU Next Generation funds. 

\section*{Data Availability}
All the data and code are publicly available from the code repository in GitHub.

\bibliographystyle{mnras}
\bibliography{spright} 

\begin{thebibliography}{}
\makeatletter
\relax
\def\mn@urlcharsother{\let\do\@makeother \do\$\do\&\do\#\do\^\do\_\do\%\do\~}
\def\mn@doi{\begingroup\mn@urlcharsother \@ifnextchar [ {\mn@doi@}
  {\mn@doi@[]}}
\def\mn@doi@[#1]#2{\def\@tempa{#1}\ifx\@tempa\@empty \href
  {http://dx.doi.org/#2} {doi:#2}\else \href {http://dx.doi.org/#2} {#1}\fi
  \endgroup}
\def\mn@eprint#1#2{\mn@eprint@#1:#2::\@nil}
\def\mn@eprint@arXiv#1{\href {http://arxiv.org/abs/#1} {{\tt arXiv:#1}}}
\def\mn@eprint@dblp#1{\href {http://dblp.uni-trier.de/rec/bibtex/#1.xml}
  {dblp:#1}}
\def\mn@eprint@#1:#2:#3:#4\@nil{\def\@tempa {#1}\def\@tempb {#2}\def\@tempc
  {#3}\ifx \@tempc \@empty \let \@tempc \@tempb \let \@tempb \@tempa \fi \ifx
  \@tempb \@empty \def\@tempb {arXiv}\fi \@ifundefined
  {mn@eprint@\@tempb}{\@tempb:\@tempc}{\expandafter \expandafter \csname
  mn@eprint@\@tempb\endcsname \expandafter{\@tempc}}}

\bibitem[\protect\citeauthoryear{{Acu{\~n}a}, {Lopez}, {Morel}, {Deleuil},
  {Mousis}, {Aguichine}, {Marcq}  \& {Santerne}}{{Acu{\~n}a}
  et~al.}{2022}]{Acuña2022}
{Acu{\~n}a} L.,  {Lopez} T.~A.,  {Morel} T.,  {Deleuil} M.,  {Mousis} O.,
  {Aguichine} A.,  {Marcq} E.,   {Santerne} A.,  2022, \mn@doi [\aap]
  {10.1051/0004-6361/202142374}, \href
  {https://ui.adsabs.harvard.edu/abs/2022A&A...660A.102A} {660, A102}

\bibitem[\protect\citeauthoryear{{Aguichine}, {Mousis}, {Deleuil}  \&
  {Marcq}}{{Aguichine} et~al.}{2021}]{Aguichine21}
{Aguichine} A.,  {Mousis} O.,  {Deleuil} M.,   {Marcq} E.,  2021, \mn@doi
  [\apj] {10.3847/1538-4357/abfa99}, \href
  {https://ui.adsabs.harvard.edu/abs/2021ApJ...914...84A} {914, 84}

\bibitem[\protect\citeauthoryear{{Bashi}, {Helled}, {Zucker}  \&
  {Mordasini}}{{Bashi} et~al.}{2017}]{Bashi17}
{Bashi} D.,  {Helled} R.,  {Zucker} S.,   {Mordasini} C.,  2017, \mn@doi [\aap]
  {10.1051/0004-6361/201629922}, \href
  {https://ui.adsabs.harvard.edu/abs/2017A&A...604A..83B} {604, A83}

\bibitem[\protect\citeauthoryear{{Batalha} et~al.,}{{Batalha}
  et~al.}{2013}]{Batalha13}
{Batalha} N.~M.,  et~al., 2013, \mn@doi [\apjs] {10.1088/0067-0049/204/2/24},
  \href {https://ui.adsabs.harvard.edu/abs/2013ApJS..204...24B} {204, 24}

\bibitem[\protect\citeauthoryear{{Bean}, {Raymond}  \& {Owen}}{{Bean}
  et~al.}{2021}]{Bean2021}
{Bean} J.~L.,  {Raymond} S.~N.,   {Owen} J.~E.,  2021, \mn@doi [Journal of
  Geophysical Research (Planets)] {10.1029/2020JE006639}, \href
  {https://ui.adsabs.harvard.edu/abs/2021JGRE..12606639B} {126, e06639}

\bibitem[\protect\citeauthoryear{{Bitsch}, {Raymond}  \& {Izidoro}}{{Bitsch}
  et~al.}{2019}]{Bitsch2019}
{Bitsch} B.,  {Raymond} S.~N.,   {Izidoro} A.,  2019, \mn@doi [\aap]
  {10.1051/0004-6361/201935007}, \href
  {https://ui.adsabs.harvard.edu/abs/2019A&A...624A.109B} {624, A109}

\bibitem[\protect\citeauthoryear{{Bluhm} et~al.,}{{Bluhm}
  et~al.}{2021}]{Bluhm2021}
{Bluhm} P.,  et~al., 2021, \mn@doi [\aap] {10.1051/0004-6361/202140688}, \href
  {https://ui.adsabs.harvard.edu/abs/2021A&A...650A..78B} {650, A78}

\bibitem[\protect\citeauthoryear{{Burn}, {Schlecker}, {Mordasini},
  {Emsenhuber}, {Alibert}, {Henning}, {Klahr}  \& {Benz}}{{Burn}
  et~al.}{2021}]{Burn21}
{Burn} R.,  {Schlecker} M.,  {Mordasini} C.,  {Emsenhuber} A.,  {Alibert} Y.,
  {Henning} T.,  {Klahr} H.,   {Benz} W.,  2021, \mn@doi [\aap]
  {10.1051/0004-6361/202140390}, \href
  {https://ui.adsabs.harvard.edu/abs/2021A&A...656A..72B} {656, A72}

\bibitem[\protect\citeauthoryear{{Chen} \& {Kipping}}{{Chen} \&
  {Kipping}}{2017}]{ChenKipping17}
{Chen} J.,  {Kipping} D.,  2017, \mn@doi [\apj] {10.3847/1538-4357/834/1/17},
  \href {https://ui.adsabs.harvard.edu/abs/2017ApJ...834...17C} {834, 17}

\bibitem[\protect\citeauthoryear{{Diamond-Lowe} et~al.,}{{Diamond-Lowe}
  et~al.}{2022}]{DiamondLowe2022}
{Diamond-Lowe} H.,  et~al., 2022, \mn@doi [\aj] {10.3847/1538-3881/ac7807},
  \href {https://ui.adsabs.harvard.edu/abs/2022AJ....164..172D} {164, 172}

\bibitem[\protect\citeauthoryear{{Dorn} \& {Lichtenberg}}{{Dorn} \&
  {Lichtenberg}}{2021}]{DornLichtenberg2021}
{Dorn} C.,  {Lichtenberg} T.,  2021, \mn@doi [\apjl]
  {10.3847/2041-8213/ac33af}, \href
  {https://ui.adsabs.harvard.edu/abs/2021ApJ...922L...4D} {922, L4}

\bibitem[\protect\citeauthoryear{{Dorn}, {Khan}, {Heng}, {Connolly}, {Alibert},
  {Benz}  \& {Tackley}}{{Dorn} et~al.}{2015}]{Dorn15}
{Dorn} C.,  {Khan} A.,  {Heng} K.,  {Connolly} J. A.~D.,  {Alibert} Y.,  {Benz}
  W.,   {Tackley} P.,  2015, \mn@doi [\aap] {10.1051/0004-6361/201424915},
  \href {https://ui.adsabs.harvard.edu/abs/2015A&A...577A..83D} {577, A83}

\bibitem[\protect\citeauthoryear{Feroz, Hobson  \& Bridges}{Feroz
  et~al.}{2009}]{Feroz2009}
Feroz F.,  Hobson M.~P.,   Bridges M.,  2009, \mn@doi [Monthly Notices of the
  Royal Astronomical Society] {10.1111/j.1365-2966.2009.14548.x}, 398, 1601

\bibitem[\protect\citeauthoryear{{Foreman-Mackey}, Hogg, Lang  \&
  Goodman}{{Foreman-Mackey} et~al.}{2013}]{Foreman-Mackey2012}
{Foreman-Mackey} D.,  Hogg D.~W.,  Lang D.,   Goodman J.,  2013, \mn@doi
  [Publications of the Astronomical Society of the Pacific] {10.1086/670067},
  125, 306

\bibitem[\protect\citeauthoryear{{Fulton} et~al.,}{{Fulton}
  et~al.}{2017}]{Fulton17}
{Fulton} B.~J.,  et~al., 2017, \mn@doi [\aj] {10.3847/1538-3881/aa80eb}, \href
  {https://ui.adsabs.harvard.edu/abs/2017AJ....154..109F} {154, 109}

\bibitem[\protect\citeauthoryear{Gelman, Carlin, Stern, Dunson, Vehtari  \&
  Rubin}{Gelman et~al.}{2013}]{Gelman2013}
Gelman A.,  Carlin J.~B.,  Stern H.~S.,  Dunson D.~B.,  Vehtari A.,   Rubin
  D.~B.,  2013, Bayesian {{Data Analysis}}.
{CRC Press}, \mn@doi{10.1201/b16018}

\bibitem[\protect\citeauthoryear{{Ginzburg}, {Schlichting}  \&
  {Sari}}{{Ginzburg} et~al.}{2018}]{Ginzburg2018}
{Ginzburg} S.,  {Schlichting} H.~E.,   {Sari} R.,  2018, \mn@doi [\mnras]
  {10.1093/mnras/sty290}, \href
  {https://ui.adsabs.harvard.edu/abs/2018MNRAS.476..759G} {476, 759}

\bibitem[\protect\citeauthoryear{{Hatzes} \& {Rauer}}{{Hatzes} \&
  {Rauer}}{2015}]{HatzesRauer15}
{Hatzes} A.~P.,  {Rauer} H.,  2015, \mn@doi [\apjl]
  {10.1088/2041-8205/810/2/L25}, \href
  {https://ui.adsabs.harvard.edu/abs/2015ApJ...810L..25H} {810, L25}

\bibitem[\protect\citeauthoryear{Kanodia, Wolfgang, Stefansson, Ning  \&
  Mahadevan}{Kanodia et~al.}{2019}]{Kanodia2019}
Kanodia S.,  Wolfgang A.,  Stefansson G.~K.,  Ning B.,   Mahadevan S.,  2019,
  \mn@doi [The Astrophysical Journal] {10.3847/1538-4357/ab334c}, 882, 38

\bibitem[\protect\citeauthoryear{Kass \& Raftery}{Kass \&
  Raftery}{1995}]{Kass1995}
Kass R.~E.,  Raftery A.~E.,  1995, Journal of the american statistical
  association, 90, 773

\bibitem[\protect\citeauthoryear{Koposov et~al.,}{Koposov
  et~al.}{2023}]{Koposov2023}
Koposov S.,  et~al., 2023, Joshspeagle/Dynesty: V2.1.2, Zenodo,
  \mn@doi{10.5281/zenodo.7995596}

\bibitem[\protect\citeauthoryear{{Lee} \& {Chiang}}{{Lee} \&
  {Chiang}}{2016}]{LeeChiang2016}
{Lee} E.~J.,  {Chiang} E.,  2016, \mn@doi [\apj] {10.3847/0004-637X/817/2/90},
  \href {https://ui.adsabs.harvard.edu/abs/2016ApJ...817...90L} {817, 90}

\bibitem[\protect\citeauthoryear{{L{\'e}ger} et~al.,}{{L{\'e}ger}
  et~al.}{2004}]{Leger2004}
{L{\'e}ger} A.,  et~al., 2004, \mn@doi [\icarus]
  {10.1016/j.icarus.2004.01.001}, \href
  {https://ui.adsabs.harvard.edu/abs/2004Icar..169..499L} {169, 499}

\bibitem[\protect\citeauthoryear{{Lissauer} et~al.,}{{Lissauer}
  et~al.}{2011}]{Lissauer11}
{Lissauer} J.~J.,  et~al., 2011, \mn@doi [\apjs] {10.1088/0067-0049/197/1/8},
  \href {https://ui.adsabs.harvard.edu/abs/2011ApJS..197....8L} {197, 8}

\bibitem[\protect\citeauthoryear{Lopez \& Fortney}{Lopez \&
  Fortney}{2014}]{Lopez2014}
Lopez E.~D.,  Fortney J.~J.,  2014, \mn@doi [The Astrophysical Journal]
  {10.1088/0004-637X/792/1/1}, 792, 1

\bibitem[\protect\citeauthoryear{Luque \& Pall{\'e}}{Luque \&
  Pall{\'e}}{2022}]{Luque2022}
Luque R.,  Pall{\'e} E.,  2022, \mn@doi [Science] {10.1126/science.abl7164},
  377, 1211

\bibitem[\protect\citeauthoryear{Mills \& Mazeh}{Mills \&
  Mazeh}{2017}]{Mills2017}
Mills S.~M.,  Mazeh T.,  2017, \mn@doi [The Astrophysical Journal]
  {10.3847/2041-8213/aa67eb}, 839, L8

\bibitem[\protect\citeauthoryear{{Mousis}, {Deleuil}, {Aguichine}, {Marcq},
  {Naar}, {Aguirre}, {Brugger}  \& {Gon{\c{c}}alves}}{{Mousis}
  et~al.}{2020}]{Mousis20}
{Mousis} O.,  {Deleuil} M.,  {Aguichine} A.,  {Marcq} E.,  {Naar} J.,
  {Aguirre} L.~A.,  {Brugger} B.,   {Gon{\c{c}}alves} T.,  2020, \mn@doi
  [\apjl] {10.3847/2041-8213/ab9530}, \href
  {https://ui.adsabs.harvard.edu/abs/2020ApJ...896L..22M} {896, L22}

\bibitem[\protect\citeauthoryear{{Neil}, {Liston}  \& {Rogers}}{{Neil}
  et~al.}{2022}]{Neil2022}
{Neil} A.~R.,  {Liston} J.,   {Rogers} L.~A.,  2022, \mn@doi [\apj]
  {10.3847/1538-4357/ac609b}, \href
  {https://ui.adsabs.harvard.edu/abs/2022ApJ...933...63N} {933, 63}

\bibitem[\protect\citeauthoryear{Ning, Wolfgang  \& Ghosh}{Ning
  et~al.}{2018}]{Ning2018}
Ning B.,  Wolfgang A.,   Ghosh S.,  2018, \mn@doi [The Astrophysical Journal]
  {10.3847/1538-4357/aaeb31}, 869, 5

\bibitem[\protect\citeauthoryear{{Otegi}, {Bouchy}  \& {Helled}}{{Otegi}
  et~al.}{2020}]{Otegi20}
{Otegi} J.~F.,  {Bouchy} F.,   {Helled} R.,  2020, \mn@doi [\aap]
  {10.1051/0004-6361/201936482}, \href
  {https://ui.adsabs.harvard.edu/abs/2020A&A...634A..43O} {634, A43}

\bibitem[\protect\citeauthoryear{{Owen} \& {Wu}}{{Owen} \&
  {Wu}}{2017}]{OwenWu2017}
{Owen} J.~E.,  {Wu} Y.,  2017, \mn@doi [\apj] {10.3847/1538-4357/aa890a}, \href
  {https://ui.adsabs.harvard.edu/abs/2017ApJ...847...29O} {847, 29}

\bibitem[\protect\citeauthoryear{Parviainen}{Parviainen}{2015}]{Parviainen2015}
Parviainen H.,  2015, \mn@doi [Monthly Notices of the Royal Astronomical
  Society] {10.1093/mnras/stv894}, 450, 3233

\bibitem[\protect\citeauthoryear{Parviainen}{Parviainen}{2018}]{Parviainen2018}
Parviainen H.,  2018, in , Handbook of {{Exoplanets}}.
{Springer International Publishing}, {Cham}, pp 1--24 (\mn@eprint {arxiv}
  {1711.03329}), \mn@doi{10.1007/978-3-319-30648-3_149-1}

\bibitem[\protect\citeauthoryear{{Piaulet} et~al.,}{{Piaulet}
  et~al.}{2023}]{Piaulet2023}
{Piaulet} C.,  et~al., 2023, \mn@doi [Nature Astronomy]
  {10.1038/s41550-022-01835-4}, \href
  {https://ui.adsabs.harvard.edu/abs/2023NatAs...7..206P} {7, 206}

\bibitem[\protect\citeauthoryear{Price, Storn  \& Lampinen}{Price
  et~al.}{2005}]{Price2005}
Price K.,  Storn R.,   Lampinen J.,  2005, Differential {{Evolution}}.
{Springer}, {Berlin}, \mn@doi{10.1007/978-0-387-36896-2}

\bibitem[\protect\citeauthoryear{{Ricker} et~al.,}{{Ricker}
  et~al.}{2014}]{TESS}
{Ricker} G.~R.,  et~al., 2014, in {Oschmann} Jacobus~M. J.,  {Clampin} M.,
  {Fazio} G.~G.,   {MacEwen} H.~A.,  eds,  Society of Photo-Optical
  Instrumentation Engineers (SPIE) Conference Series Vol. 9143, Space
  Telescopes and Instrumentation 2014: Optical, Infrared, and Millimeter Wave.
  p. 914320 (\mn@eprint {arXiv} {1406.0151}), \mn@doi{10.1117/12.2063489}

\bibitem[\protect\citeauthoryear{Robert}{Robert}{2007}]{Robert2007}
Robert C.~P.,  2007, The {{Bayesian Choice}}.
{Springer}, {New York}

\bibitem[\protect\citeauthoryear{{Rogers}}{{Rogers}}{2015}]{Rogers15}
{Rogers} L.~A.,  2015, \mn@doi [\apj] {10.1088/0004-637X/801/1/41}, \href
  {https://ui.adsabs.harvard.edu/abs/2015ApJ...801...41R} {801, 41}

\bibitem[\protect\citeauthoryear{{Rogers}, {Schlichting}  \& {Owen}}{{Rogers}
  et~al.}{2023}]{Rogers2023}
{Rogers} J.~G.,  {Schlichting} H.~E.,   {Owen} J.~E.,  2023, \mn@doi [\apjl]
  {10.3847/2041-8213/acc86f}, \href
  {https://ui.adsabs.harvard.edu/abs/2023ApJ...947L..19R} {947, L19}

\bibitem[\protect\citeauthoryear{{Schlecker} et~al.,}{{Schlecker}
  et~al.}{2022}]{Schlecker22}
{Schlecker} M.,  et~al., 2022, \mn@doi [\aap] {10.1051/0004-6361/202142543},
  \href {https://ui.adsabs.harvard.edu/abs/2022A&A...664A.180S} {664, A180}

\bibitem[\protect\citeauthoryear{Skilling}{Skilling}{2004}]{Skilling2004}
Skilling J.,  2004, \mn@doi [AIP Conference Proceedings] {10.1063/1.1835238},
  735, 395

\bibitem[\protect\citeauthoryear{Skilling}{Skilling}{2006}]{Skilling2006}
Skilling J.,  2006, in {{ISBA}} 8th {{World Meeting}} on {{Bayesian
  Statistics}}.

\bibitem[\protect\citeauthoryear{{Southworth}}{{Southworth}}{2011}]{tepcat}
{Southworth} J.,  2011, \mn@doi [\mnras] {10.1111/j.1365-2966.2011.19399.x},
  \href {https://ui.adsabs.harvard.edu/abs/2011MNRAS.417.2166S} {417, 2166}

\bibitem[\protect\citeauthoryear{Speagle}{Speagle}{2020}]{Speagle2020}
Speagle J.~S.,  2020, \mn@doi [Monthly Notices of the Royal Astronomical
  Society] {10.1093/mnras/staa278}, 493, 3132

\bibitem[\protect\citeauthoryear{{Spergel} et~al.,}{{Spergel}
  et~al.}{2015}]{Spergel2015}
{Spergel} D.,  et~al., 2015, \mn@doi [arXiv e-prints]
  {10.48550/arXiv.1503.03757}, \href
  {https://ui.adsabs.harvard.edu/abs/2015arXiv150303757S} {p. arXiv:1503.03757}

\bibitem[\protect\citeauthoryear{{Venturini}, {Guilera}, {Haldemann}, {Ronco}
  \& {Mordasini}}{{Venturini} et~al.}{2020}]{Venturini2020}
{Venturini} J.,  {Guilera} O.~M.,  {Haldemann} J.,  {Ronco} M.~P.,
  {Mordasini} C.,  2020, \mn@doi [\aap] {10.1051/0004-6361/202039141}, \href
  {https://ui.adsabs.harvard.edu/abs/2020A&A...643L...1V} {643, L1}

\bibitem[\protect\citeauthoryear{Weiss \& Marcy}{Weiss \&
  Marcy}{2014}]{Weiss2014}
Weiss L.~M.,  Marcy G.~W.,  2014, \mn@doi [The Astrophysical Journal]
  {10.1088/2041-8205/783/1/L6}, 783, L6

\bibitem[\protect\citeauthoryear{Wolfgang, Rogers  \& Ford}{Wolfgang
  et~al.}{2016}]{Wolfgang2016}
Wolfgang A.,  Rogers L.~A.,   Ford E.~B.,  2016, \mn@doi [The Astrophysical
  Journal] {10.3847/0004-637X/825/1/19}, 825, 19

\bibitem[\protect\citeauthoryear{Zeng et~al.,}{Zeng et~al.}{2019}]{Zeng2019}
Zeng L.,  et~al., 2019, \mn@doi [Proceedings of the National Academy of
  Sciences] {10.1073/pnas.1812905116}, 116, 9723

\makeatother
\end{thebibliography}

\appendix

\section{Parameter posteriors}

\begin{figure*}
    \centering
    \includegraphics[width=\columnwidth]{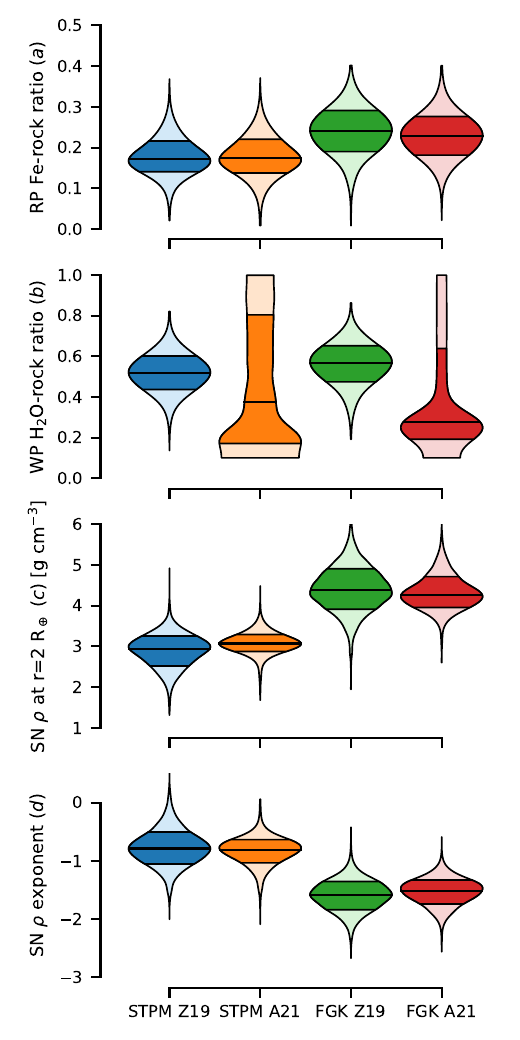}
    \includegraphics[width=\columnwidth]{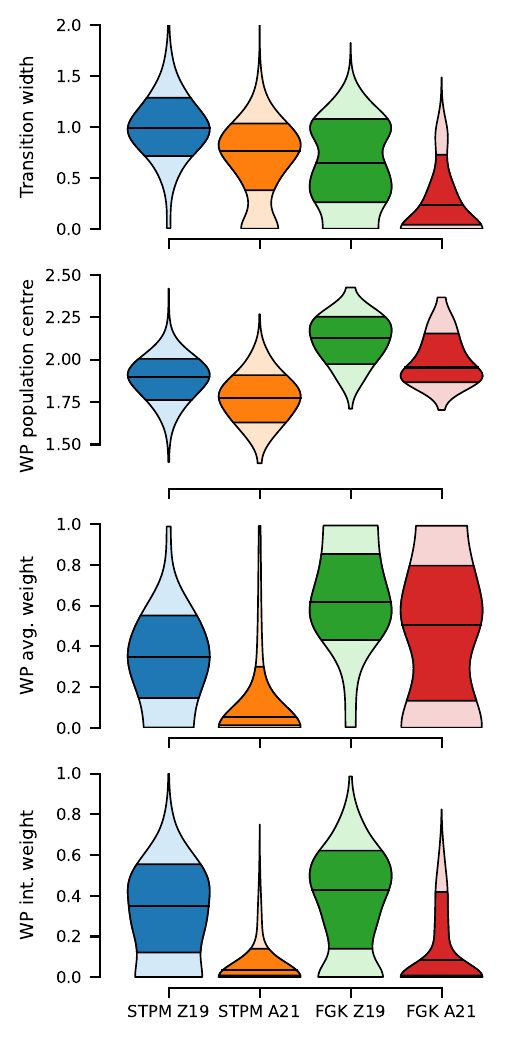}
    \caption{Posterior distributions for the analytical mixture model parameters inferred from the updated STPM catalogue and TEPCat FGK~star host catalogue, and quantities derived from the model parameters. RP refers to rocky planets, WP to water-rich planets, and SN to hydrogen-rich sub-Neptunes, while Z19 refers to the \citet{Zeng2019} water-rich planet density models, and A21 to the \citet{Aguichine21} water-rich planet density models. The transition width is the width of the transition region between rocky planets and sub-Neptunes ($r_4 - r_1$), WP population centre is the centre-of-mass radius for the water world population calculated from the water world population weights over the transition region, WP average weight is the average water world population weight over the transition region, and WP integrated weight is the weight of the water world population integrated over all planetary radii.}
    \label{fig:parameter_posteriors}
\end{figure*}

\begin{figure*}
    \centering
    \includegraphics[width=\textwidth]{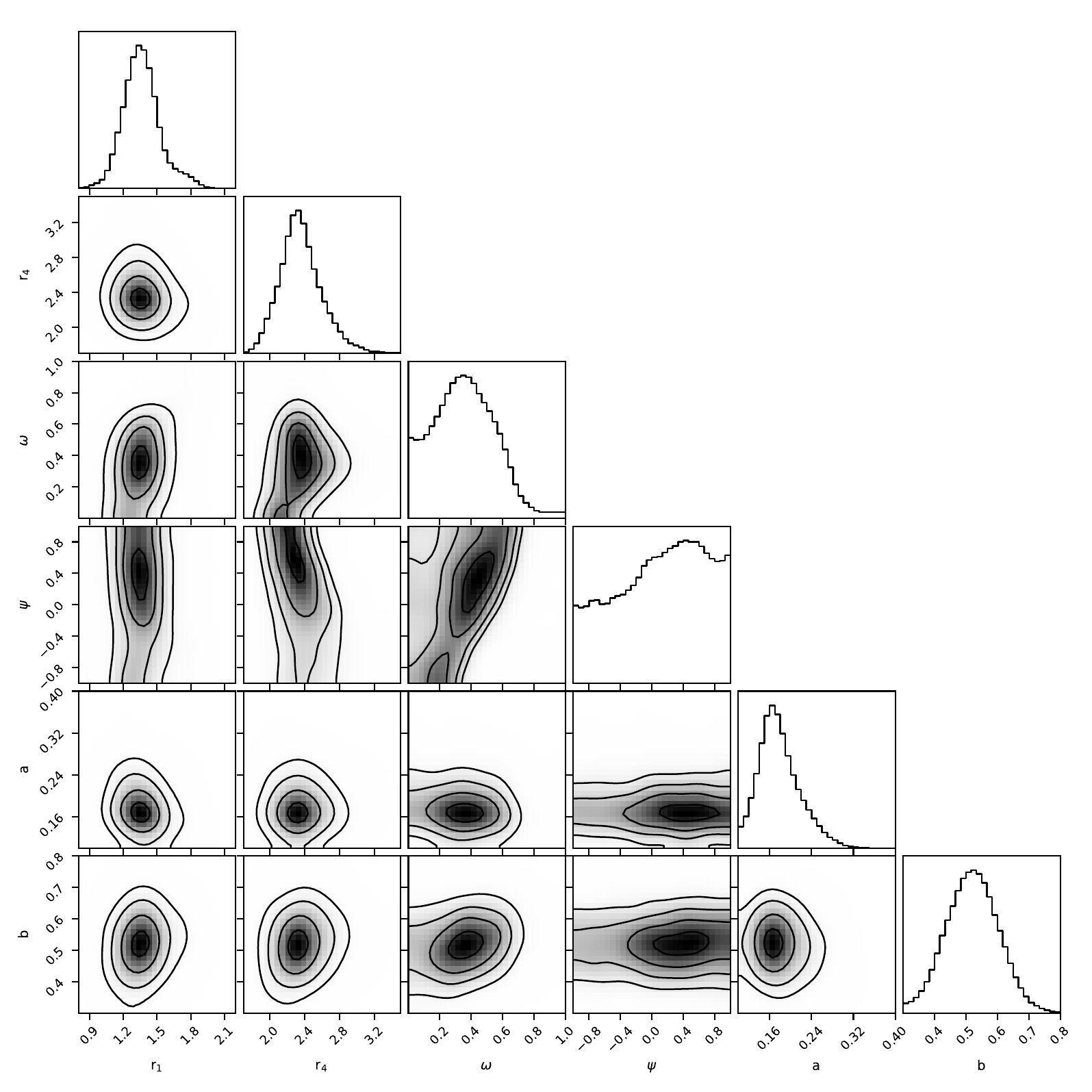}
    \caption{Joint posterior distributions for a subset of the analytical mixture model parameters inferred from the updated STPM catalogue using the \citet{Zeng2019} water-rich planet density models.}
    \label{fig:parameter_joint_posteriors_stpm}
\end{figure*}

\begin{figure*}
    \centering
    \includegraphics[width=\textwidth]{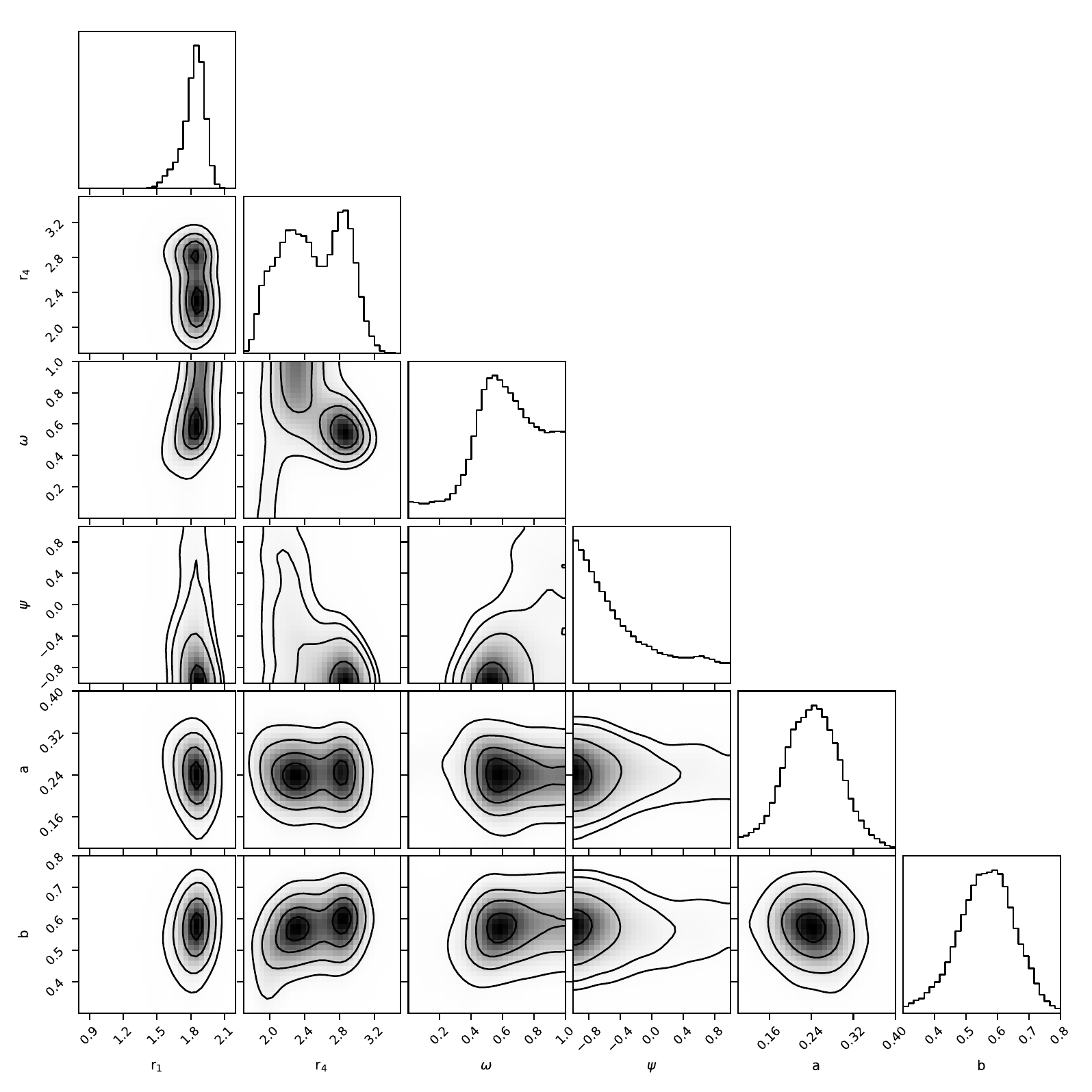}
    \caption{Joint posterior distributions for a subset of the analytical mixture model parameters inferred from the TEPCat FGK catalogue using the \citet{Zeng2019} water-rich planet density models.}
    \label{fig:parameter_joint_posteriors_fgk}
\end{figure*}

\begin{figure*}
    \centering
    \includegraphics[width=\textwidth]{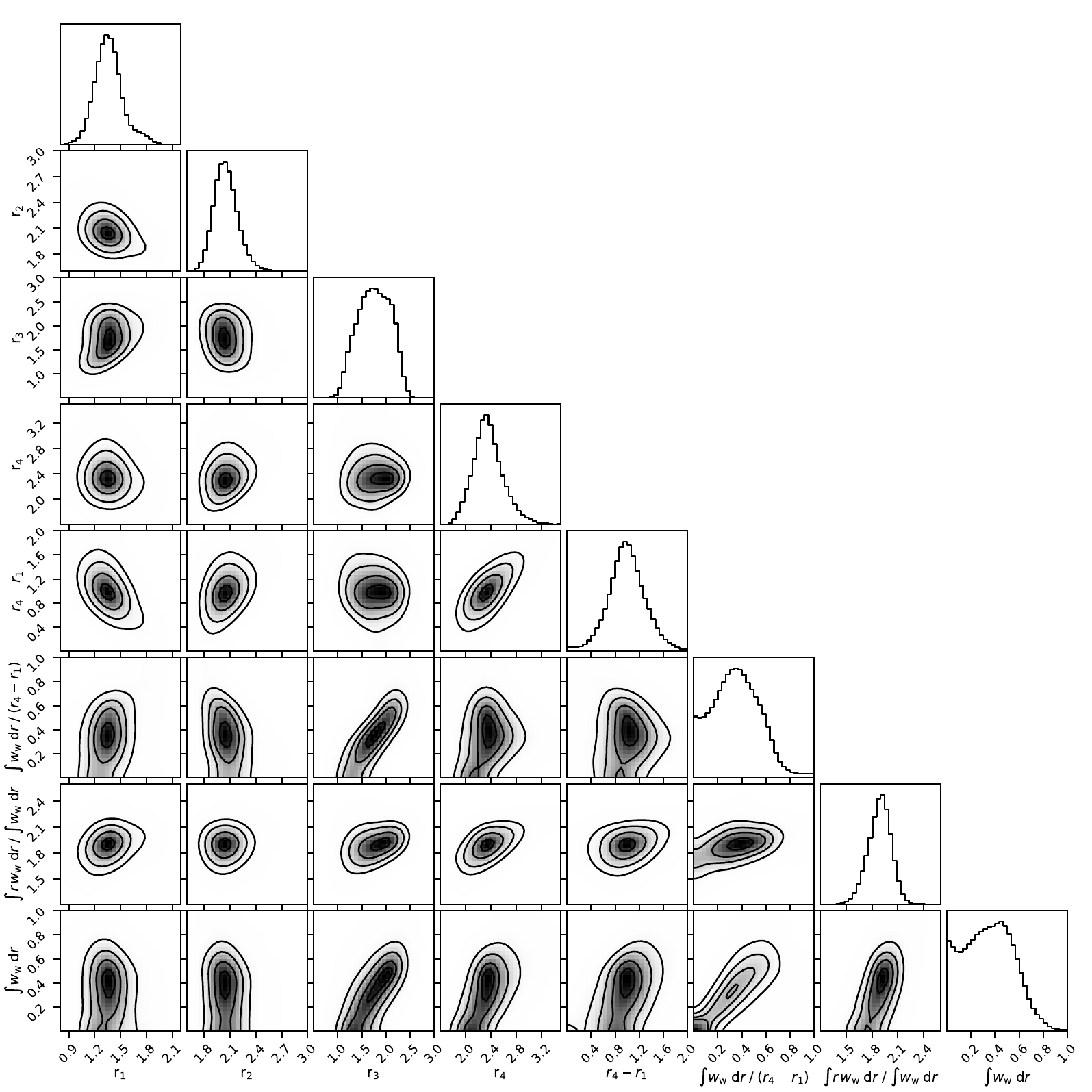}
    \caption{Joint posterior distributions for a set of analytical mixture model parameters and quantities derived from the model parameters inferred from the updated STPM catalogue using the \citet{Zeng2019} water-rich planet density models. Here, $r_4-r_1$ is the width of the transition region between rocky planets and sub-Neptunes, $\left(\int w_\mathrm{w}(r) \mathrm{d}r\right) / \left(r_4 - r_1\right)$ is the mean water world population weight over the transition region,  $\left(\int r w_\mathrm{w} \mathrm{d}r \right)/\left(\int w_\mathrm{w}(r) \mathrm{d}r\right)$ is the water world population centre calculated as a weighted mean of the planet radius, and  $\int w_\mathrm{w}(r) \mathrm{d}r$ is the total integrated water world population weight.}
    \label{fig:derived_parameter_joint_posteriors_stpm}
\end{figure*}

\begin{figure*}
    \centering
    \includegraphics[width=\textwidth]{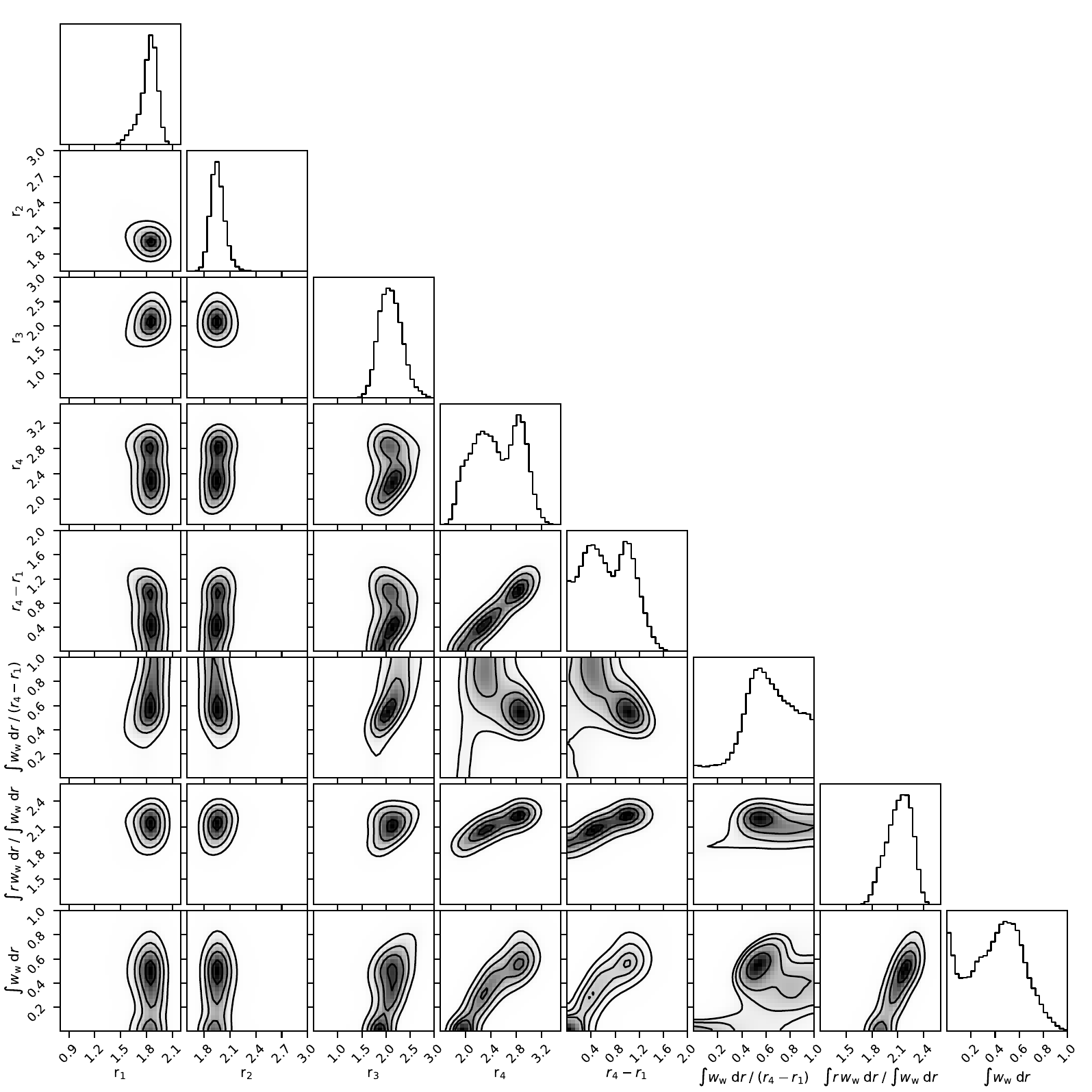}
    \caption{As in Fig.~\ref{fig:parameter_joint_posteriors_stpm} but for the TEPCat FGK host star sample.}
    \label{fig:derived_parameter_joint_posteriors_fgk}
\end{figure*}

\begin{figure*}
    \centering
    \includegraphics[width=\textwidth]{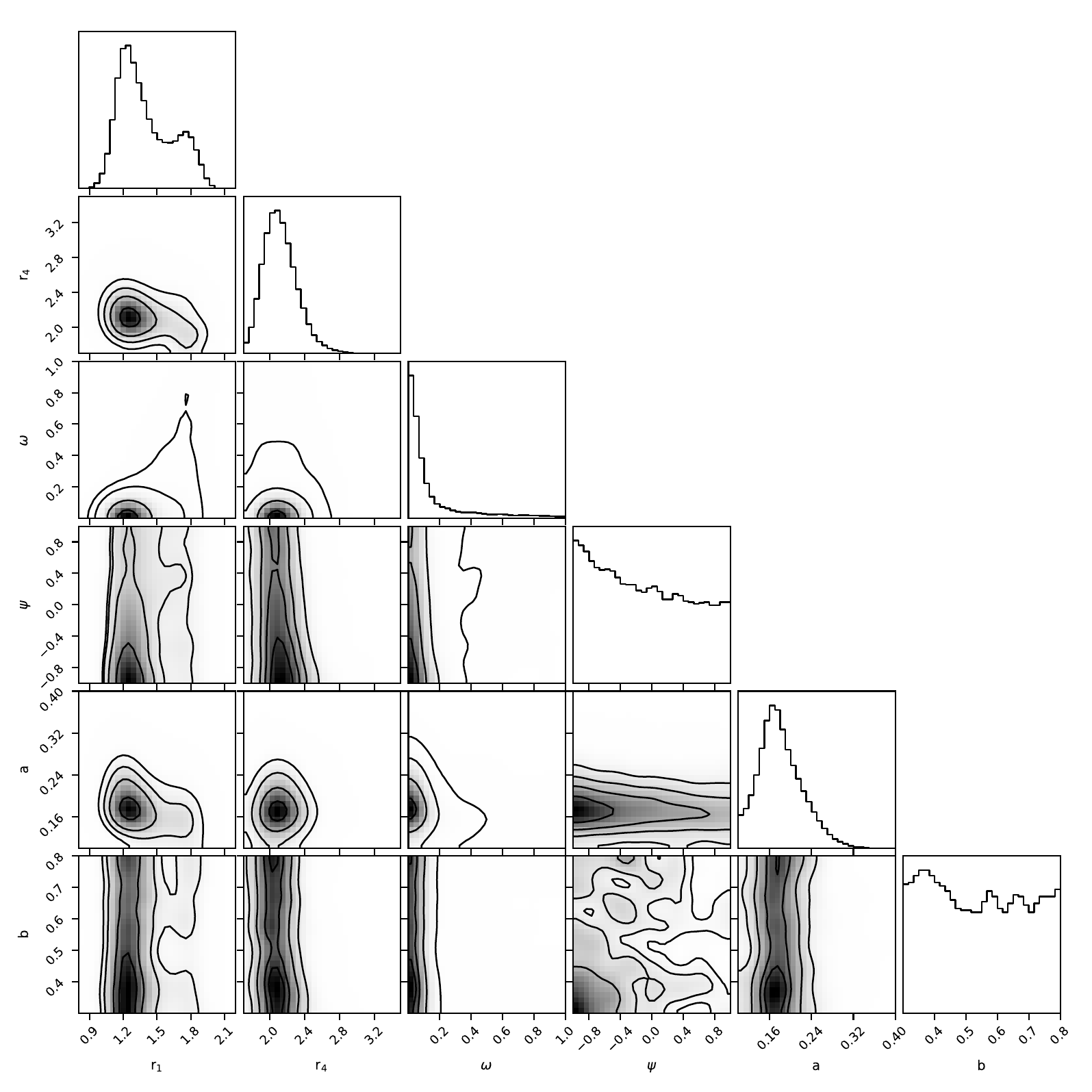}
    \caption{Joint posterior distributions for a subset of the analytical mixture model parameters inferred from the updated STPM catalogue using the \citet{Aguichine21} water-rich planet density models.}
    \label{fig:parameter_joint_posteriors_stpm_a21}
\end{figure*}

\begin{figure*}
    \centering
    \includegraphics[width=\textwidth]{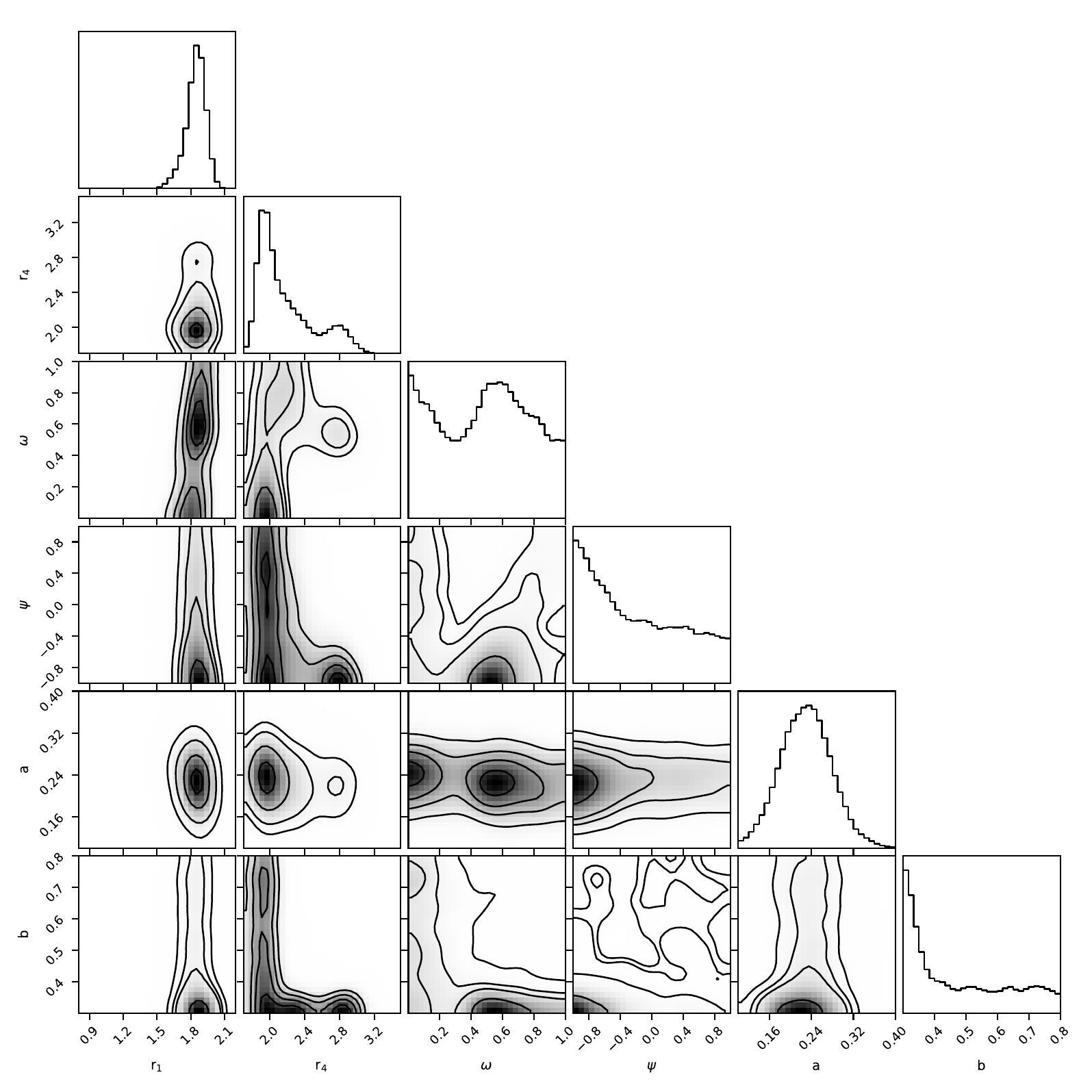}
    \caption{Joint posterior distributions for a subset of the analytical mixture model parameters inferred from the TEPCat FGK catalogue using the \citet{Aguichine21} water-rich planet density models.}
    \label{fig:parameter_joint_posteriors_fgk_a21}
\end{figure*}

\begin{figure*}
    \centering
    \includegraphics[width=\textwidth]{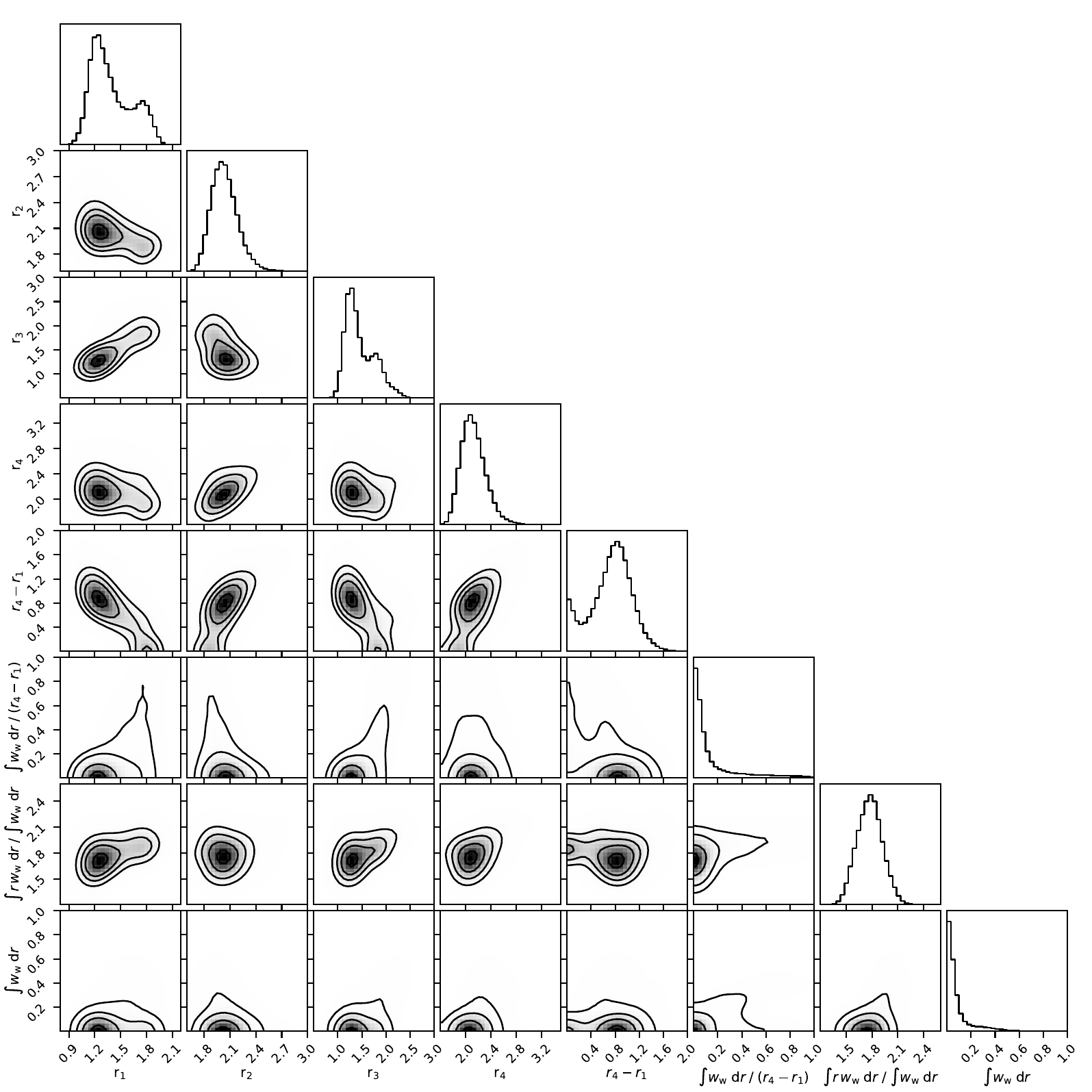}
    \caption{Joint posterior distributions for a set of analytical mixture model parameters and quantities derived from the model parameters inferred from the updated STPM catalogue using the \citet{Aguichine21} water-rich planet density models. Here, $r_4-r_1$ is the width of the transition region between rocky planets and sub-Neptunes, $\left(\int w_\mathrm{w}(r) \mathrm{d}r\right) / \left(r_4 - r_1\right)$ is the mean water world population weight over the transition region,  $\left(\int r w_\mathrm{w} \mathrm{d}r \right)/\left(\int w_\mathrm{w}(r) \mathrm{d}r\right)$ is the water world population centre calculated as a weighted mean of the planet radius, and  $\int w_\mathrm{w}(r) \mathrm{d}r$ is the total integrated water world population weight.}
    \label{fig:derived_parameter_joint_posteriors_stpm_a21}
\end{figure*}

\begin{figure*}
    \centering
    \includegraphics[width=\textwidth]{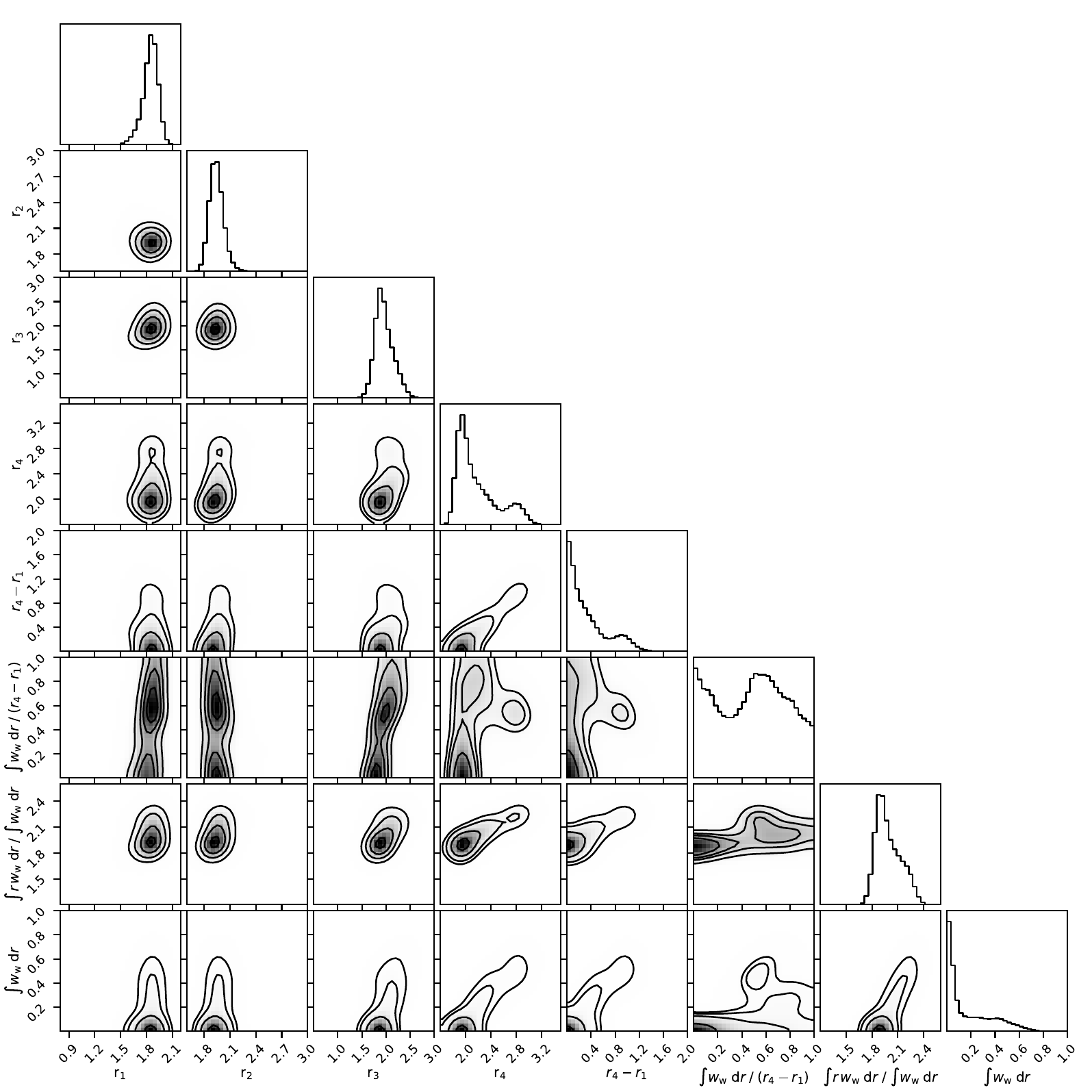}
    \caption{As in Fig.~\ref{fig:parameter_joint_posteriors_stpm_a21} but for the TEPCat FGK host star sample.}
    \label{fig:derived_parameter_joint_posteriors_fgk_a21}
\end{figure*}

\bsp	
\label{lastpage}
\end{document}